\documentclass[letterpaper]{article} 
\usepackage{aaai25}  
\usepackage{times}  
\usepackage{helvet}  
\usepackage{courier}  
\usepackage[hyphens]{url}  
\usepackage{graphicx} 
\usepackage{natbib}  
\usepackage{caption} 
\usepackage[ruled,vlined]{algorithm2e}
\SetKwInput{KwInput}{Input}                
\SetKwInput{KwOutput}{Output}   

\newtheorem{definition}{Definition}

\newtheorem{proposition}{Proposition}

\newtheorem{Observation}{Observation}
\usepackage{subfigure}

\newcommand{\NE}{\mathbf{\bar{x}}}

\newcommand{\R}{\mathbb{R}}

\usepackage{amsmath}
\usepackage{amsfonts}

\title{Asymptotic Extinction in Large Coordination Games}
\author{Desmond Chan \textsuperscript{\rm 1}
\footnote{Corresponding author}, 
Bart de Keijzer \textsuperscript{\rm 1}, 
Tobias Galla \textsuperscript{\rm 2}, 
Stefanos Leonardos \textsuperscript{\rm 1},
Carmine Ventre \textsuperscript{\rm 1}}
\affiliations{
    \textsuperscript{\rm 1} King's College London, 
     \textsuperscript{\rm 2} Institute for Cross-Disciplinary Physics and Complex Systems (IFISC, CSIC-UIB)\\
     desmond.chan@kcl.ac.uk , bart.de$\_$keijzer@kcl.ac.uk, tobias.galla@ifisc.uib-csic.es, \\
     stefanos.leonardos@kcl.ac.uk , carmine.ventre@kcl.ac.uk
}
\renewcommand{\vec}[1]{\mathbf{#1}}

\begin{document}

\maketitle
\begin{abstract}
We study the exploration-exploitation trade-off for large multiplayer coordination games where players strategise via Q-Learning, a common learning framework in multi-agent reinforcement learning. Q-Learning is known to have two shortcomings, namely  non-convergence and potential equilibrium selection problems, when there are multiple fixed points, called Quantal Response Equilibria (QRE). Furthermore, whilst QRE have full support for finite games, it is not clear how Q-Learning behaves as the game becomes large. 
In this paper, we characterise the critical exploration rate that guarantees convergence to a unique fixed point, addressing the two shortcomings above. Using a generating-functional method, we show that this rate increases with the number of players and the alignment of their payoffs. For many-player coordination games with perfectly aligned payoffs, this exploration rate is roughly twice that of $p$-player zero-sum games. As for large games, we provide a structural result for QRE, which suggests that as the game size increases, Q-Learning converges to a QRE near the boundary of the simplex of the action space, a phenomenon we term asymptotic extinction, where a constant fraction of the actions are played with zero probability at a rate $o(1/N)$ for an $N$-action game.
\end{abstract}

\section{Introduction}
Multi-agent systems are an increasingly relevant area in AI 
research. They typically consist of 
learning agents 
trying to coordinate to reach specific outcomes, such scenarios are prevalent in fields ranging from  economics \citep{march1991exploration}, robotics and distributed systems \citep{panait2005cooperative}. A key challenge in these settings is balancing 
exploration and exploitation in high-dimensional action spaces. 
 Exploration is required for the discovery of optimal strategies; this can come at the expense of short-term rewards. Effectively exploring such complex spaces can be a critical point of failure, preventing convergence to ``good'' outcomes. 
 
 In multi-agent reinforcement learning (MARL),   coordination scenarios consisting of interacting agents, can be represented as games.  Throughout this work, we will focus on Q-Learning, one of the most widely used methods in MARL, as it provides a framework to analyse the exploitation-exploration trade-off algorithmically.  The fixed points of Q-Learning  are Quantal Response Equilibria (QRE) \citep{leonardos2021exploration, leonardos2022exploration}, which always assign positive probability to all actions of finite games and for low exploration rates approximate the Nash Equilibria (NE) for the underlying game. 

Coordination games are characterised by players' payoffs being aligned in a manner to incentivise picking mutually beneficial actions. In such settings, agents following the Q-Learning algorithm over a fixed game can exhibit two different dynamical behaviours: (i) Convergence to a unique fixed point (at high exploration rates) -- where agents reach the same, joint fixed point regardless of initial conditions; and, (ii) Convergence to multiple equilibria (at low exploration rates) -- the final strategy profiles agents converge to is dependent on initial conditions. The effectiveness of Q-Learning  is influenced by which of these two outcomes emerges during the learning process.  
This paper investigates the dynamical behaviour of Q-Learning over large, multi-player coordination games where the payoff matrices are randomly drawn from multivariate Gaussians.  In each game, the payoffs matrices are randomly-generated and held fixed.  Players are 
assigned random initial strategies and we study the emerging dynamics. 


\noindent \textbf{Related Work and Our Contribution.}
The study of random competitive games was considered in \citep{galla2013complex}, which was inspired by replicator models in the context of biological evolution \citep{opper1992phase, Galla_2006}.
Following this line of work, we characterise, through the use of random games, the typical behaviour in complex, coordination games a priori the learning process. Our work complements and extends the results in \citep{sanders2018prevalence} to coordination games.

Our theoretical analysis suggest in coordination games with large action sets of size $N$,  a constant, non-zero proportion of actions are played with a frequency of $o(1/N)$.  We call this effect \emph{asymptotic extinction}. This extinction rate is asymptotic and varies with the model parameters. While simulations cannot fully quantitatively confirm this effect, simulation results are broadly consistent with our theoretical analysis. 
 Taking this effect into account, a minimum exploration rate $T_{\text{crit}}$, which guarantees convergence to a unique fixed point can be found over games with varying number of players and degree of payoff correlation.

\section{Preliminaries}
\paragraph*{Multiplayer normal form games.}
A $p$ player, $N$ action normal form game, $\mathcal{G}$, is defined by a tuple, $\mathcal{G} = (\mathcal{P}, \mathcal{A}, \Pi) $, where $\mathcal{P} :=\{1, 2, \dots, p\}$ is the \emph{set of players}, and $\mathcal{A}:= \{ 1, \ldots, N\}$ is a set of actions. Each player in $\mathcal{G}$ chooses an action, resulting in an \emph{action profile}, i.e., an element $\vec{a} \in \mathcal{A}^p$. Thus, for an action profile $\vec{a}$, we write $a_i$ to refer to Player $i$'s chosen action in $\vec{a}$. Furthermore, we use $\vec{a}_{-i}$ to refer to vector obtained from $\vec{a}$ by removing the $i$th coordinate. The notation $(b, \vec{a}_{-i})$ then refers to the vector obtained from $\vec{a}$ by replacing the value at coordinate $i$ with $b$ (so that $\vec{a} = (a_i,\vec{a}_{-i})$.
For each action profile, every player experiences a certain \emph{payoff} which players want to maximise. Payoffs are specified by the \emph{payoff function} 
$\Pi : \mathcal{A}^p \rightarrow \mathbb{R}^p$, where for $i \in \mathcal{P}$ and $\vec{a} \in \mathcal{A}^p$, the payoff for Player $i$ on action profile $\vec{a}$ is given by $\Pi(\vec{a})_i$. 

Players can choose their actions probabilistically. This gives rise to the notion of a \emph{strategy} $\vec{x}$, which is a  probability distribution over $\mathcal{A}$. Thus, $\vec{x}$ is a point on the $(N-1)$-simplex $\Delta_N = \{ \mathbf{x} \in \mathbb{R}^n_{\geq 0}\ :\ x_1 + \cdots + x_N = 1 \}$. 
\emph{Interior points} of the simplex correspond to strategies where all actions are played with positive probability $(x_i > 0, \forall i)$. Points not in the interior are known as \emph{boundary points}. Similar to the notion of an action profile, a \emph{strategy profile} is a choice of strategy by each of the players, and is hence given by an element $\vec{x} = (\vec{x}^1, \ldots, \vec{x}^p) \in \Delta_N^p$. A strategy profile $\vec{x}$, induces a probability distribution over action profiles, and we define the payoff $R(\vec{x})^i$ of Player $i$ for $\vec{x}$ from $\Pi$, as the expected value of the payoff of the random action profile:
\begin{equation}\label{eq:payoff}
R(\vec{x})^i = \sum_{\vec{a} \in \mathcal{A}^p}
\Pi(\vec{a})_i \prod_{i \in \mathcal{P}} x^i_{a_i}.
\end{equation}
Similar to our notation for action profiles, for a strategy profile $\vec{x}$ we use $\vec{x}^{-i}$ to refer to the vector obtained from $\vec{x}$ by removing the strategy of Player $i$ from it. The notation $(y, \vec{x}^{-i})$ then refers to the vector obtained from $\vec{x}$ by replacing $\vec{x}^i$ with $y \in \Delta_N$. Furthermore, we sometimes abuse notation and write $(a,\vec{x}^{-i})$, for an action $a \in \mathcal{A}$, to denote $(\vec{e}_a, \vec{x}^{-i})$, where $\vec{e}_a$ denotes the vector with a $1$ at coordinate $a$ and $0$s at all other coordinates.

\noindent \textbf{Constructing payoff matrices.}
To generate a game, we draw the payoff matrix, $\Pi$, from a multivariate Gaussian with mean $0$ which treats all players symmetrically \footnote{The choice of Gaussian distribution can be motivated by a maximum entropy and universality  argument \cite{tao2011random} See the Appendix for more details.}.The covariance matrix of the distribution is determined by parameter $\Gamma \in (-1, p-1)$, which captures the \emph{pairwise correlations} between the players' payoffs.  Fixing all but two players $i,j \in \mathcal{P}$, we have the following pairwise-correlation structure for each action profile $\vec{a, b } \in \mathcal{A}^p$.
\begin{align*}
    \mathbb{E}\left[
    \Pi(\vec{a})_{i} \cdot 
    \Pi(\vec{b})_{j}
    \right]
    = 
\begin{cases}
1, & \text{ if } \vec{a} = \vec{b}, i = j , \\
\Gamma/(p-1), & \text{ if }  \vec{a} = \vec{b}, i \neq j \\
0, & \text{ if } \vec{a} \neq \vec{b}
\end{cases}
\end{align*}
$\Gamma$ acts as a measure of the level of cooperativeness-competitiveness of a game. For every additional unit of reward that Player $i$ receives by changing their strategy, the sum of all other players' payoffs will change by $\Gamma$ in expectation. $\Gamma= -1$ represents a $p$-player zero-sum game, while $\Gamma=p-1$ represents an identical payoff game. 
In general, $\Gamma < 0$, corresponds to \emph{competitive games} where players can only benefit at the expense of others. Conversely, $\Gamma > 0$, corresponds to \emph{coordination games} in which the player's payoffs are positively aligned to a degree given by $\Gamma$. 

\noindent \textbf{Q-Learning.} 
Given a game and an initial set of strategies, we wish to analyse how players learn and how their strategies evolve over time. Players following a learning algorithm turn games into dynamical systems with strategies evolving in a state space. Our focus is on the Q-Learning model \citep{watkins1992q}.
\footnote{The use of Q-Learning is widespread in multi-agent learning and game theory literature where it appears under various names and variants including Experience Weighted Attraction (EWA) \citep{camerer1999experience}, Boltzmann Q-Learning \citep{kianercy2012dynamics, bloembergen2015evolutionary}
etc. See \citep{pangallo2017towards} for a general overview.} Here each player $i \in \mathcal{P}$ keeps track of a Q-value corresponding to each action $a \in \mathcal{A}$, which estimates the \emph{quality} of the given action. At each time step $t$, the Q-value corresponding to action $a$ are updated as follows:
\begin{align}\label{eq:qlearning}
    Q_a^i(t+1) 
    &= 
    \underbrace{(1- \alpha) Q_a^i(t)}_{\text{discounted previous Q-value}}
    + 
    \underbrace{R(a, \mathbf{x}^{-i}(t))^i}_{\text{current reward}}
\end{align}
where $\alpha \in (0,1)$ denotes the discount parameter. \footnote{Note that we could include a factor $\alpha$ in front of the reward term, this should extend the possible range in which one can trade off exploration with exploitation.}
The discount rate is then given by $(1-\alpha)$ which indicates experience (the previous Q-value) is prioritised against the current reward. With the Q-values, player select mixed strategies according to the softmax distribution parameterised by $\beta>0$
\begin{align}\label{eq:softmax}
    x_a^i(t) & =\frac{\exp \left[\beta Q_a^i(t)\right]}{\sum_{b \in \mathcal{A}} \exp \left[\beta Q_b^i(t)\right]} 
\end{align}
 We  refer to parameter $T:= \alpha / \beta $ as the \emph{exploration rate}.  Taking $\alpha, \beta \to 0$, but keeping $T$ constant is equivalent to taking smaller step sizes in each update until we reach the continuous limit.  See the Appendix for details on how this limit is obtained from the discrete equations \eqref{eq:qlearning} \eqref{eq:softmax}.
Thus, we obtain the continuous Q-Learning equations \cite{sato2003coupled, tuyls:qlearning}:
\begin{align}
    \frac{\dot{ x}_a^i(t)}{x_a^i(t)}= R(a, \mathbf{x}^{-i}(t))^i - T \ln x_{a}^i(t) - \rho^i(t)  
    \label{eq:cql} 
\end{align}
where  $\rho^i:= R(\vec{x}^i(t), \mathbf{x}^{-i}(t))^i - T \langle \vec{x}^i(t), \ln \vec{x}^i(t) \rangle$ is a normalisation parameter, which ensures strategies stay within the simplex. Here, $\langle\cdot, \cdot \rangle$ denotes the inner product. The fixed points of Q-Learning dynamics (both discrete and continuous variant) are \emph{Quantal Response Equilibria}.
\begin{definition}[Quantal Response Equilibrium (QRE)] 
A strategy profile $\NE \in \Delta$ is a
\emph {Quantal Response Equilibrium} (QRE) if, for all players $i \in \mathcal{P}$ and all actions $a\in\mathcal{A}$ 
    \begin{equation*}
        \bar{x}_{a}^i =
        \frac{\exp(R(a, \NE^{-i})^i/T)}
        {\sum_{j \in N}
       \exp(R(j, \NE^{-i})^i/T)}.
    \end{equation*}
    where $T \in [0, \infty)$ denotes the \emph{exploration rate}, which is assumed to be equal for all players.
\end{definition}

\noindent \textbf{QRE Interpretation}
QREs are a natural equilibrium solution concept, which takes into account the risk-reward management of the players, and are related to NE. At $T=0$, only the actions which yield the highest payoff are played. Here the QRE corresponds to the NE. For $T > 0$, players mix actions, with players converging to the uniform distribution as $T \to \infty$. Thus, $T$ acts as a \emph{risk aversion} parameter. Crucially, for any finite game, any initial strategy in the interior of the simplex, continuous Q-Learning \eqref{eq:cql} will converge to a unique interior fixed point given a sufficiently high $T$ \citep{hussain2023asymptotic}.

\noindent \textbf{Rescaling of $T$}
As we vary the number of actions, $N$, available to each player, intuitively, we expect the `typical action' ${x}_{a}^i$ to scale at a rate of $1/N$.  Hence, the expected payoff across different actions scales at a rate of $\sqrt{1/ N^{(p-1)}}$. We dedicate a segment in the Appendix to discuss this rescaling.
To facilitate a fair comparison across games of different sizes, we have to take these effects into account and rescale $T$ as follows:
$T = \tilde{T}/\sqrt{N^{(p-1)}}$, where $\tilde{T}$ represents the previous unscaled exploration rate.
Thus, we will henceforth be working with the scaled exploration rate $T$.
 
\subsubsection{Overview of Numerical Results}
For all values of $\Gamma$, Q-Learning converges to a unique fixed point at sufficiently high exploration rates $T$. Below some critical exploration rate $T_{\text{crit}}$, which increases with $\Gamma$ and $p$, we observe dynamics of varying nature, given in Table \ref{table_11}.
\begin{table}[h!]
\centering
\begin{tabular}{|c|c|}
\hline
\textbf{Condition} & \textbf{Dynamical Behaviour at $T< T_{\text{crit}}$} \\ 
\hline
$\Gamma > 0$ & Convergence to multiple fixed points \\ 
\hline
$\Gamma \approx 0$ & Occasional limit cycles \\ 
\hline
$\Gamma < 0$ & Chaotic behaviour \\ 
\hline
\end{tabular}
\caption{Dynamical behaviour at $T< T_{\text{crit}}$ for varying $\Gamma$.  A brief overview (and supporting figures from simulation) of the possible dynamical behaviours can be found in the Appendix.  A similar overview for this model can be found in \citep{sanders2018prevalence}.}
\label{table_11}
\end{table}
\subsection{Analytic Background}
We provide an overview of the generating functional method, which was first  introduced in \citep{galla2013complex} to study the dynamical behaviour of Q-Learning in the $N \to \infty$ limit. Instead of focusing on the outcome of a single initialisation of Q-Learning, we study the evolution of ensembles of possible initialisations.  Thus, we will work with the distributions of possible Q-Learning trajectories and how they evolve over time. Borrowing methods from dynamical mean-field theory (DMFT) \footnote{A step-by-step guide of the method on replicator models, we refer the reader to  \citep{galla2024generatingfunctionalanalysisrandomlotkavolterra}.}, we consider the distribution of trajectories in the $N \to \infty$ limit; here the statistics of the Q-Learning trajectories satisfy a stochastic relation, which we refer to as the \emph{effective dynamics}.
As $N$ increases, the statistics of the Q-learning trajectories obey the effective dynamics with increasing accuracy. 
 Solving for the fixed points of the effective dynamics and its corresponding stability will allow us to identify the critical exploration rate, $T_{\text{crit}}$, required for convergence to a unique fixed point and the rate of extinction in the unique fixed point regime.  

\noindent \textbf{Effective Dynamics.}
Deriving the effective dynamics relation can be broken up into the following two steps: (i) Defining a probability measure over possible trajectories under Q-Learning; (ii) Averaging over all possible payoff matrices, by considering the large action space limit ($N \to \infty$).  

The calculations in each of these steps are lengthy and relies on path integral methods from disordered systems theory and a rescaling of variables, we have relegated the details of derivation from \citep{sanders2018prevalence} into the Appendix alongside references.  The result of this analysis, which holds for any given value of $\Gamma$ and $T$, is the following effective dynamics:
\begin{align}\label{eff_dyn} 
    \frac{\dot{x}(t)}{x(t)} = & \Gamma \int_{t_0}^t G\left(t, t^{\prime}\right) C\left(t, t^{\prime}\right)^{p-2} x\left(t^{\prime}\right) \mathrm{d} t^{\prime} \notag   \\ 
    & - T \ln x(t)-\rho(t)+\eta(t).
\end{align}
The term $\rho(t)$ is a function corresponding to the normalisation term $\rho^i(t)$ of \eqref{eq:cql}, and $\eta(t)$ is a coloured (i.e., time-correlated) Gaussian random variable satisfying: 
    $\left\langle \eta(t)\eta(t^{\prime}) \right\rangle_{*} = C\left(t, t^{\prime}\right)^{p-1}$
for all $t^{\prime} < t$, where $\left\langle \dots \right\rangle_{*}$ denotes the expected value over realisations. 

The $\eta$ term can be thought of as the randomness at the fixed point phase originating from the initialisation of the payoff matrices. Lastly, $C$ and $G$ are given by:
\begin{equation*}
C\left(t, t^{\prime}\right)= \left\langle x(t) x(t')\right\rangle_{*} , \;G\left(t, t^{\prime}\right)=\left\langle\frac{\delta x(t)}{\delta \eta\left(t^{\prime}\right)}\right\rangle_{*} .
\end{equation*}
Here, $C$ describes time correlations between strategies and $G$ acts a `response' function that links how strategies are correlated over time and how this varies with $\eta$.

Equation \eqref{eff_dyn} describes the evolution of the marginal probability of playing a given action $x(t)$ as a stochastic process.  It is scaled by a factor of $N$ such that each action is played with mean $1$, $\left\langle x(t) \right\rangle_{*} = 1$.\footnote{This is discussed in more detail in the Appendix, along with the scaling of the exploration rate $T$.} We note that \eqref{eff_dyn} does not depend on the player nor the index of the actions. This a due to the a-priori symmetry among players in the initial conditions. 
By solving for the fixed-point of \eqref{eff_dyn}, we are able to find the marginal probability distributions of playing each action at the unique fixed point regime of Q-Learning. 

\noindent \textbf{Fixed Points of the Effective Dynamics.}
For high exploration rates $T$, Q-Learning converges to a unique fixed point regardless of initial conditions.  For large $N$, this should align with the presence of a \emph{stable} fixed point for  \eqref{eff_dyn}. 
Thus, to check the dynamical behaviour of Q-Learning, given $\Gamma$ and $T$,  we would have to (i) identify the fixed points corresponding to \eqref{eff_dyn} and (ii) analyse the respective stability.

Thus, to find a fixed point of the effective dynamics in a large game, we consider the following.
First, a \emph{fixed point of \eqref{eff_dyn}} is defined as a solution where $\dot{x}(t) = 0$ (for all $t$), so that $x(t) = x$ is constant across $t$ for each realisation of the random variable $\eta$. In this stationary regime, we note: $C$ becomes a constant; $C(t, t')= \left\langle (x)^2 \right\rangle_{*} = q$, and $\eta$ turns out to be a static realisation of a Gaussian $\eta \sim \mathcal{N}\left(0,\, q^{p-1}\right)$, while $G$ is now a function of the time difference; $G(t, t') = G(t-t')$. We let $z \sim \mathcal{N} (0, 1)$ such that $\eta = q^{(p-1)/2} z$.  Thus, we have:
\begin{equation}\label{fixed_point_eq}
   0 = x(z) \left[ \Gamma q^{(p-2)} x(z) \chi - T \ln {x(z)} - \rho + q^{(p-1)/2}z \right]
\end{equation}
where we note that $x$ in \eqref{fixed_point_eq} is a function of the Gaussian realisation $z$, $\chi = \int_0^{\infty} G(\tau) d \tau$ is assumed to be finite, and $\rho$ corresponds to $\rho(t)$ in \eqref{eff_dyn}, which must be constant in $t$ as well by the requirement that $\dot{x}(t) = 0$. 
By the definitions of $q, \chi, C, G$ and $\eta$, the following self-consistency relations must hold:
\begin{align}\label{self_suff}
     \left\langle\frac{\delta x(z)}{\delta z}\right\rangle_{*} = q^{(p-1)/2}  \chi ,\; 
    \left\langle x(z)^2 \right\rangle_{*} = q,
    \left\langle x(z)\right\rangle_{*} = 1 
\end{align}
where $\langle x(z)\rangle_{*}$  can be replaced by the integral $\int^{\infty}_{-\infty} x(z)\exp(-z^2/ 2)/ \sqrt{2 \pi} dz$.
This gives us a 4-equation, 4-unknown problem (with unknowns $q,\chi,\rho,x$ and Equations \eqref{self_suff}, and \eqref{fixed_point_eq}), and solving this problem yields us a fixed point corresponding to \eqref{eff_dyn}.

\section{Beyond Competitive Games}
Up to this point, we have summarised previous work by \cite{galla2013complex, sanders2018prevalence}. While the effective dynamics \eqref{eff_dyn} and fixed point equations \eqref{fixed_point_eq} hold for any $\Gamma$, only the competitive case ($\Gamma < 0$) has been studied analytically. Now, we will explore the implications of allowing $\Gamma$ to be positive.

The structure is as follows: We show that the theory suggests asymptotic extinction occurs when $\Gamma \geq 0$, and this is consistent with numerical simulations. We solve $x(z)$ for the fixed point equations \eqref{fixed_point_eq} using an \emph{ansatz} that accounts for this effect, determining the frequency of asymptotic extinction. We will then determine $T_{\text{crit}}$, which we will compare against numerical simulations. Additional details, derivations, and supporting figures are provided in the Appendix.
\subsection{Asymptotic Extinction}
To satisfy \eqref{fixed_point_eq}, we need either of the following to hold for all values of $z$:  $x(z)=0$ or the bracketed term in \eqref{fixed_point_eq} $= 0$, whilst satisfying the self-consistency relations \eqref{self_suff}.  We can classify our fixed points into two distinct cases: i) Interior , where: $\left[ \Gamma q^{(p-2)}  x(z) \chi - T \ln x(z) - \rho+ q^{(p-1)/2} z \right] = 0 , \forall z \in \R$ ii) Boundary, where $x(z) = 0$: for some $z$. 

Points on the boundary correspond to strategies with extinct actions.  
Our analysis shows that when $\Gamma > 0$, \eqref{fixed_point_eq} has no corresponding internal fixed points, only boundary fixed points, even at high exploration rates.  How could \eqref{fixed_point_eq} suggest that Q-Learning would converge to a boundary point?
\begin{figure}[t]
    \centering
    \includegraphics[width= 0.95\linewidth]{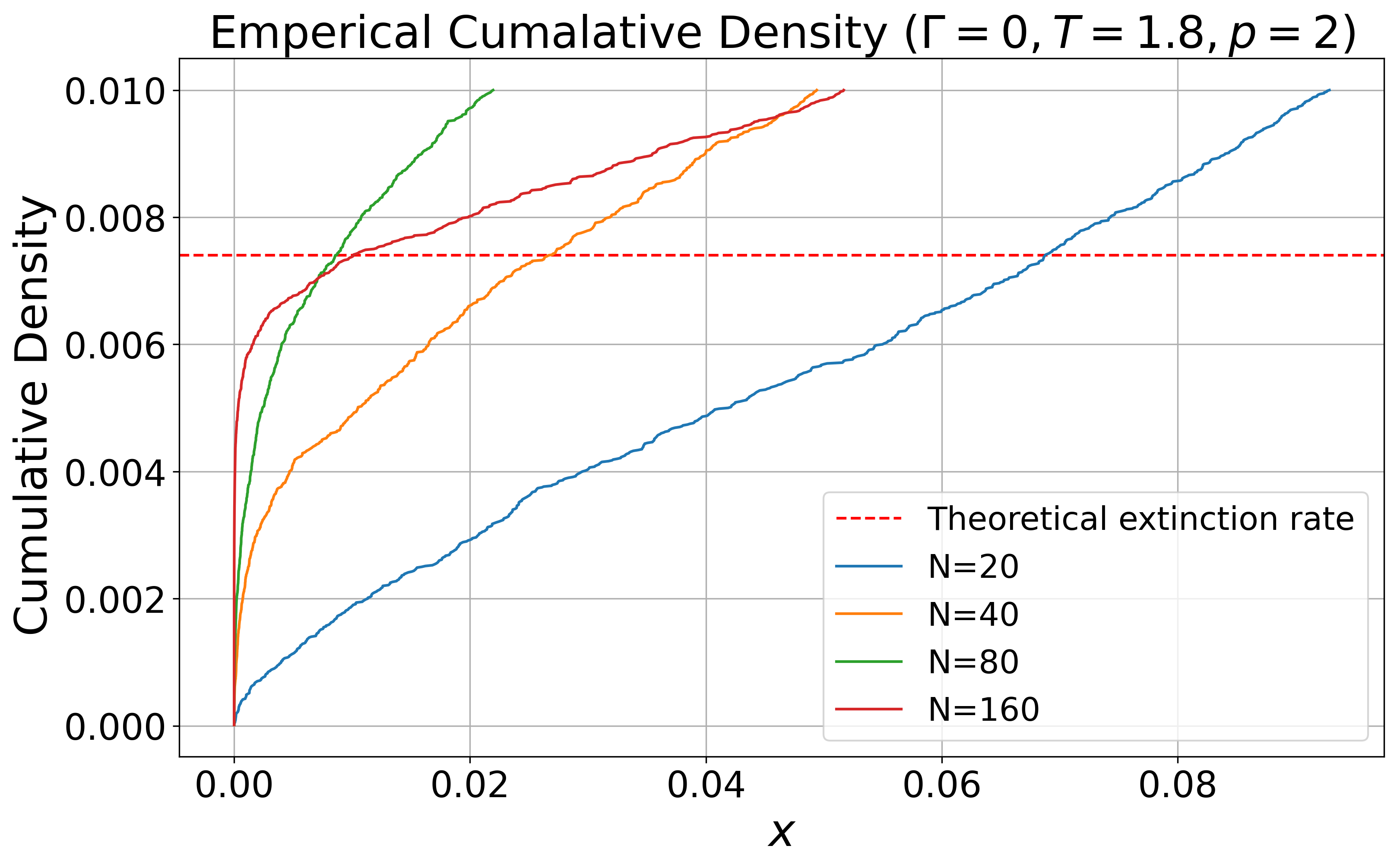}
    \caption{Empirical cumulative density plot representing the marginal likelihood of playing an action at unique fixed point for randomly generated games following the Q-Learning dynamic where $\Gamma= 0, T= 1.8, p=2$. The plot is zoomed in at the bottom $1 \%$ of least played actions and $x$ is rescaled such that $x=1$ would represent the average likelihood $(1/N)$.  As $N$ increases, a probability mass appears to form near 0, representing actions going asymptotically extinct.  The red line represents the theoretical estimate of the extinction rate $(0.74\%)$ in the $N \to \infty$ limit.  In this limit, the cumulative density plot would begin on the red line.}
    \label{fig:cdf}
\end{figure}

It is known that strategies on the boundary of the simplex are unstable under Q-Learning in any finitely sized game.  However, boundary points, as defined here, are not necessarily unstable since we are working in the large action size limit $N \to \infty$. 
(Recall $x(z)$ is rescaled by a factor of $N$ in the effective dynamics $\eqref{eff_dyn}$).  
Simulating games with fixed parameter values $(\Gamma \geq 0$ and varying $N$, we find that a proportion of actions are played with near 0 probability.  
When $T< T_{\text{crit}}$, this proportion can be significant. This occurs to a much smaller extent in the unique fixed point regime $T > T_{\text{crit}}$, typically affecting less than $1\%$ of actions. (See Figure \ref{fig:cdf}, where a probability mass of actions are played with near $0$ probability as $N$ increases.)  
This provides the basis for \emph{asymptotic extinction}, where points can be internal for any finite game, but asymptotically approaches the boundary in the large action size $N \to \infty$ limit.

The fixed point solution of $x(z)$ allows us to characterise this behaviour in the unique fixed point regime $T > T_{\text{crit}}$.
Depending on the sign of $\Gamma$, the  fixed point solution given by $x(z)$ varies significantly (See Table \ref{table:fixedpoints}).  We will provide a case-by-case ansatz of $x(z)$ \footnote{A guide to solving these relations numerically is provided in the Appendix.}, which takes the effect of asymptotic extinction into account for $\Gamma \geq 0$:
\begin{table}[h!]
\centering
\begin{tabular}{|c|c|}
\hline
\textbf{Condition} & \textbf{Stable Interior fixed point?} \\ 
\hline
$\Gamma > 0$ & no \\ 
\hline
$\Gamma = 0$ & depends on $T$ \\ 
\hline
$\Gamma < 0$ & yes \\ 
\hline
\end{tabular}
\caption{Existence condition for a stable interior fixed point for finite exploration rates.}
\label{table:fixedpoints}
\end{table}

\noindent \textbf{Competitive Games $\mathbf{(\Gamma <  0)}$.}
In this regime, there  exists a unique interior fixed point. At this fixed point we have:
\begin{align} \label{fixed_gamma_neg}
x(z)=  Ke^{bz+ax(z)}
\end{align}
where: $a = \Gamma q^{(p-2)}  \chi T^{-1} < 0 $, $b = q^{(p-1)/2} T^{-1} > 0$ and $K$ is a normalisation constant, ensuring $\langle x(z)\rangle_*=1$.  These are determined by solving the self-consistency relation \eqref{self_suff} and is in agreement with \citep{sanders2018prevalence}.
\begin{figure}[t]
    \centering
    \includegraphics[width= 0.95\linewidth]{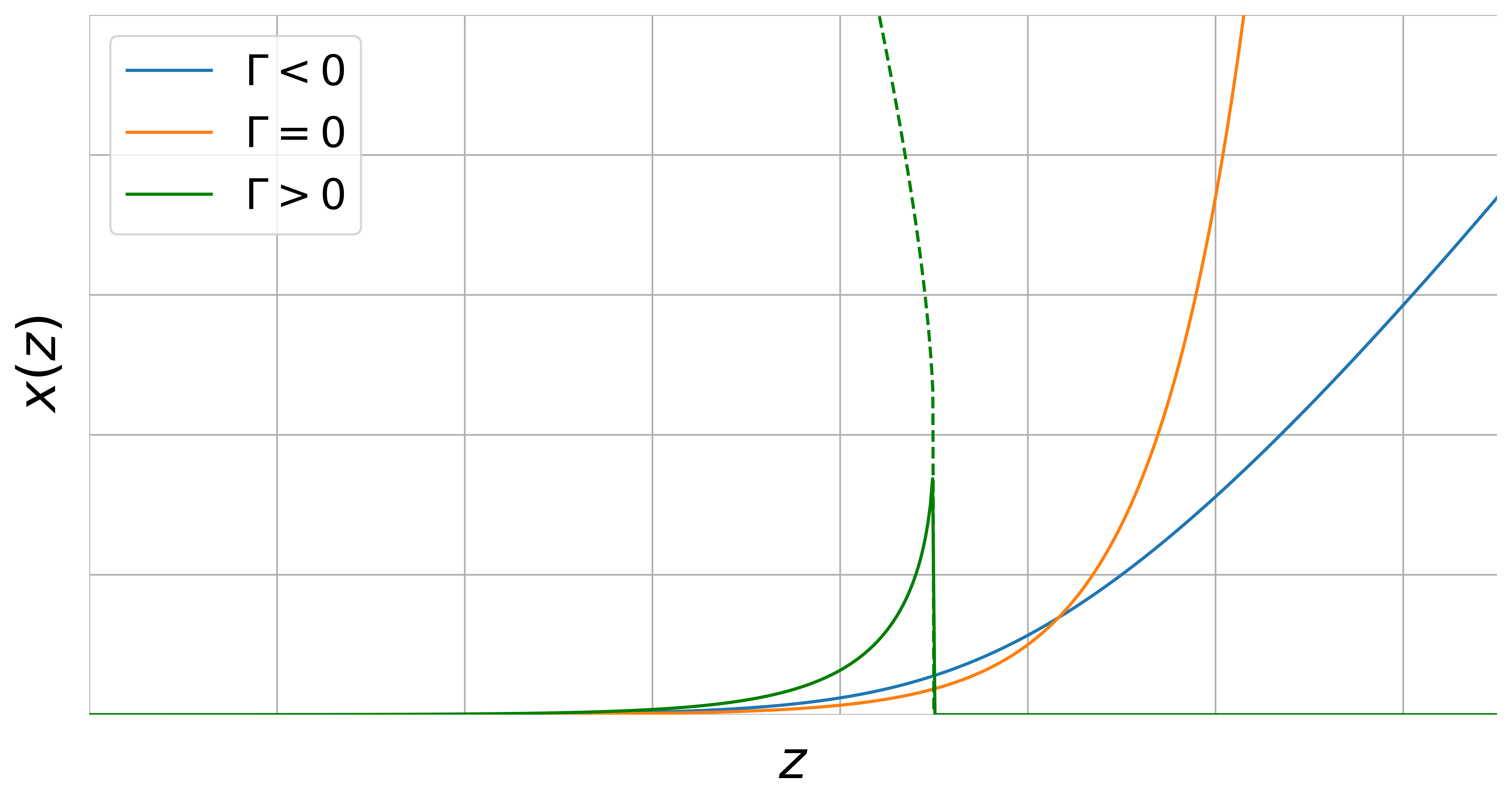}
    \caption{Sketch of $x(z)$ for different values of $\Gamma$.  The solution for $\Gamma > 0$ is double-valued below a critical $z$, as seen by the dotted lines.  The bottom (solid) branch is of interest here.}
    \label{fig:fp_dist}
\end{figure}
\noindent \textbf{Cooperative Games $\mathbf{(\Gamma > 0)}$.}
For cooperative games $\Gamma > 0$, Equation \eqref{fixed_point_eq}
does not yield an interior fixed point, but rather a boundary point. Fixed points here take the following form:
\begin{align}\label{fixed_gamma_pos}
x(z)= 
\begin{cases}
Ke^{bz+ax(z)} \; \; &, z < z_{\text{crit}} \\
0 \; \; &, z \geq z_{\text{crit}}
\end{cases}
\end{align}
where: $a = \Gamma q^{(p-2)}  \chi T^{-1} > 0$ and $z_{\text{crit}} = -1/b  \left( 1 + \ln(a K)\right)    
$. Values $a, b, K$ are the same as in \eqref{fixed_gamma_neg}, except $a$ is now positive.
We note \eqref{fixed_gamma_pos} is double-valued for $x(z)$ below $z_{\text{crit}}$.  This is represented by the two branches in Figure \ref{fig:fp_dist}: the bottom branch (the solid line) and the top branch (the dotted-line). We take the bottom branch as our value for $x(z)$. 

\noindent \textbf{Uncorrelated Games $\mathbf{(\Gamma =0)}$.}
Uncorrelated games represent a special case, where the existence of an internal fixed point is dependent on $T$. When, $T \geq  \sqrt{3e(p-1)/ 2}$, we have an internal fixed point \footnote{See the Appendix for the derivation, alongside figures demonstrating the consistency with numerical simulations.} of the form: 
\begin{align}\label{fixed_gamma_zero}
x(z) = Ke^{bz} 
\end{align}
While, when $T <  \sqrt{3e(p-1)/ 2}$, the fixed point is on the boundary, taking the form:
\begin{align}
x(z)= 
\begin{cases}
Ke^{bz} \; \; &, z < z_{\text{crit}} \\
0 \; \; &, z \geq z_{\text{crit}}
\end{cases}
\end{align}
With the disappearance of $a$,  $z_{\text{crit}}$ is to be determined directly from self-consistency \eqref{self_suff} as the third unknown. 
\begin{figure*}
    \centering
    \includegraphics[width= 0.9\textwidth]{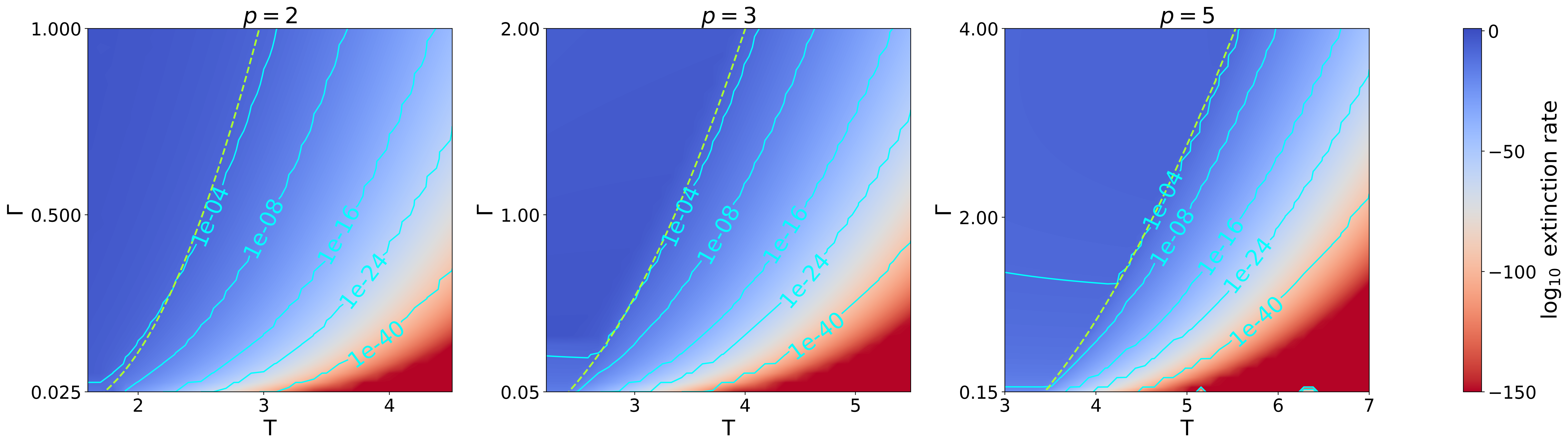}
    \caption{Theoretical asymptotic extinction rate for varying numbers of players $p$ obtained from estimates from the fixed point relations \eqref{fixed_gamma_pos}.  These estimations are only for the unique fixed point regime (right of the yellow dotted line representing the stability boundary, which is solved in the next segment) .  There are some numerical instability in the estimations (namely when $\Gamma < 0.1$, thus the axes not starting at 0), but the figure roughly shows the scale of the expected extinction rate for varying $T$ away from the boundary.}
    \label{fig:ex_rate}
\end{figure*}
\subsubsection{How likely is extinction?}
For lower exploration rates $T < T_{\text{crit}}$, the solutions of $x(z)$ are unstable and we are unable to characterise likelihood of  extinction. Our solution $x(z)$ to the fixed point relations given by \eqref{fixed_gamma_pos} and \eqref{fixed_gamma_zero} can only predict the distribution of actions when $T > T_{\text{crit}}$ (i.e. when there is a unique fixed point), with $T_{\text{crit}}$ and the corresponding regime will be identified in the next section.  For now, we will discuss the extinction likelihood obtained by solving $x(z)$, given by $P(z < z_{\text{crit}})$, where $z \sim \mathcal{N}(0,1)$. 

Figure \ref{fig:ex_rate} displays the theoretical extinction rate for games with $p \in \{2,3,5\}$ and varying $T$ and $\Gamma$. As $T$ is increased beyond $T_{\text{crit}}$, our fixed point relations suggests the likelihood of a randomly chosen strategy going extinct asymptotically decreases drastically. In the large game limit, $N \to \infty$, for any finite exploration rate $T$, theory suggests that a non-zero proportion of strategies is expected to go asymptotically extinct in coordination games.

Around the stability boundary, extinctions occur to around $1 \%$ to $0.01 \%$ of actions.  Checking selected parameter combinations of $T$ and $\Gamma$ on the stability boundary for games with more players, we find this roughly holds true for higher values of $p$.
Away from the boundary, the probability of a randomly selected action going extinct becomes very rare.  
Verifying the likelihood of asymptotic extinction experimentally in this parameter range with experiments is difficult, as extinctions become extraordinarily rare events.

\subsection{Stability Analysis}
Having found the fixed points distributions, we have to check their corresponding stability to determine if Q-Learning converges to it. We show the following result:
\begin{proposition}
Q-Learning converges to a unique fixed point when the parameters and corresponding fixed point fulfils the following relation:
\begin{align}\label{contra2}\phi
\left\langle \bigg|
\frac{ T}{ x(z)} - \Gamma 
q^{p-2} \chi
\bigg|^{-2}\right\rangle_{*} < \left((p-1) 
q^{p-2} \right)^{-1}
\end{align}
where $\phi$ is the proportion of non-extinct strategies, given by $P(z < z_{\text{crit}})$ where $z \sim \mathcal{N}(0,1)$. 
\end{proposition}
When $\Gamma < 0$, $\phi = 1$ (as the fixed points are internal); we have the same relation as \citep{sanders2018prevalence}. What we have done here is added a $\phi$-term, which takes into effect when $\Gamma \geq 0$.
\footnote{The inclusion of $\phi$ when species go extinct is standard for DMFT analysis of replicator models, see \citep{opper1992phase, galla2018dynamically}. The possibility of including a $\phi$-term for 2-player setup has been discussed in \citep{galla2013complex}, but not explored in further detail due to the apparent contradiction between theory suggesting actions going extinct, but no actions going to 0 in finite, numerical simulations.} 

Thus, \eqref{contra2} extends the analysis to the coordination setting $\Gamma \geq 0$.  In the unique fixed point regime, only a small fraction ($ < 1 \%$) of strategies go extinct, thus the relevant $\phi$s are almost always close to $1$. 
\begin{figure}[t]
    \centering
    \includegraphics[width= 0.9\linewidth]{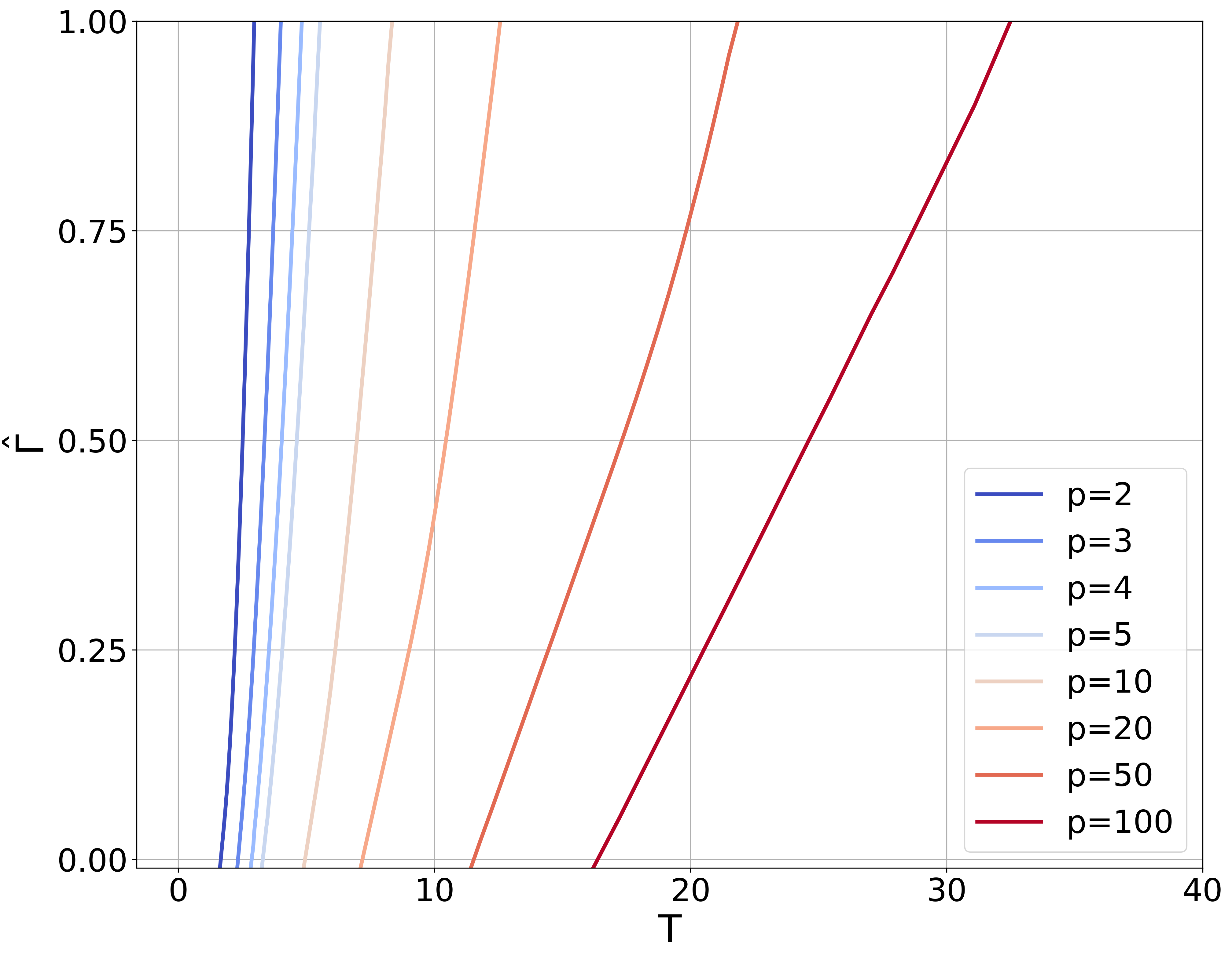}
    \caption{Stability boundary obtained by solving \eqref{contra2} for varying values of $p$, as a function of $T$ for $\hat{\Gamma} > 0 $, where $\hat{\Gamma} = \Gamma / (p-1)$.  To the right of the boundary, all Q-Learning trajectories converge to a unique fixed point in the large action size limit, $N\to \infty$.  When $\hat{\Gamma} < 0$, we recover the results from \citep{sanders2018prevalence}.  Our work extends the stability boundary to cover $\hat{\Gamma} > 0$. }
    \label{fig:stab_curve}
\end{figure}
\begin{figure}[h]
    \centering
    \includegraphics[width= 0.9\linewidth]{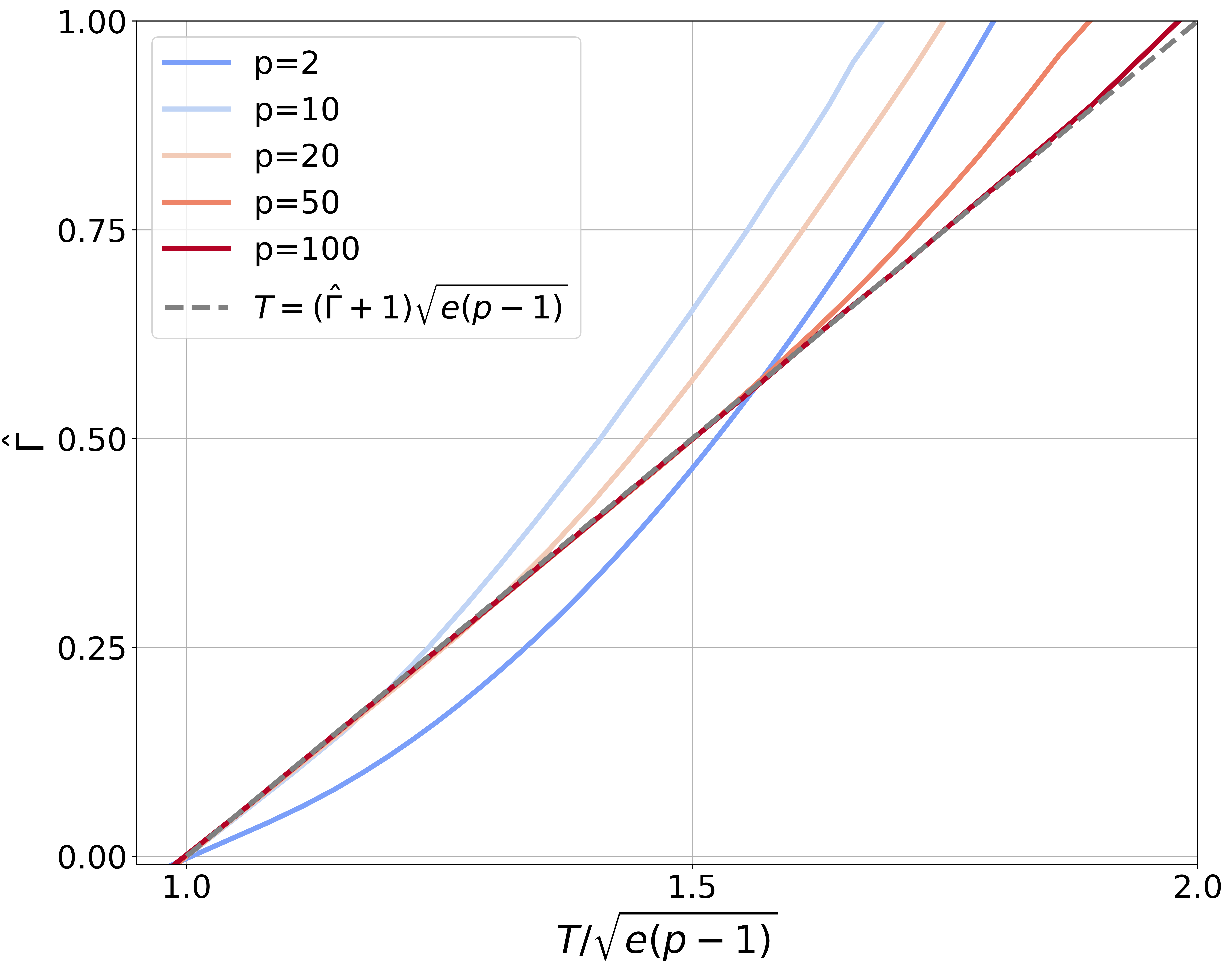}
    \caption{Rescaled stability curves for selected values of $p$.  The exploration rate, $T$, is rescaled by a factor of $\sqrt{e(p-1)}$ and the grey-dashed line represents the straight line given by $T_{\text{crit}} = (\hat{\Gamma} + 1) \sqrt{e(p-1)}$, which appears to be the limiting behaviour at $p \to \infty$. The increasing agreement with the grey line for large curves with larger values of $p$ suggests this linear relationship is valid, in the large $p$ limit.}
    \label{fig:scaled_stab_curve}
\end{figure}
More detailed guidance on obtaining \eqref{contra2} is attached in the Appendix; it is obtained by a somewhat standard procedure \footnote{See \citep{drazin2004hydrodynamic} for a textbook introduction and application of this method to fluid systems.} used to determine the linearised stability in dynamical systems, as follows: i) linearising the dynamics at the fixed point ii) taking a frequency transform and identifying a stability criterion which guarantees the stability of the whole system under all possible perturbation modes\footnote{There is a subtle deviation in the derivation of the stability relation from \citep{sanders2018prevalence} and other similar work \citep{galla2013complex}. We dedicate a short segment discussing this in the Appendix.}.
\begin{figure*}[t]
    \centering
    \includegraphics[width= 0.9\textwidth]{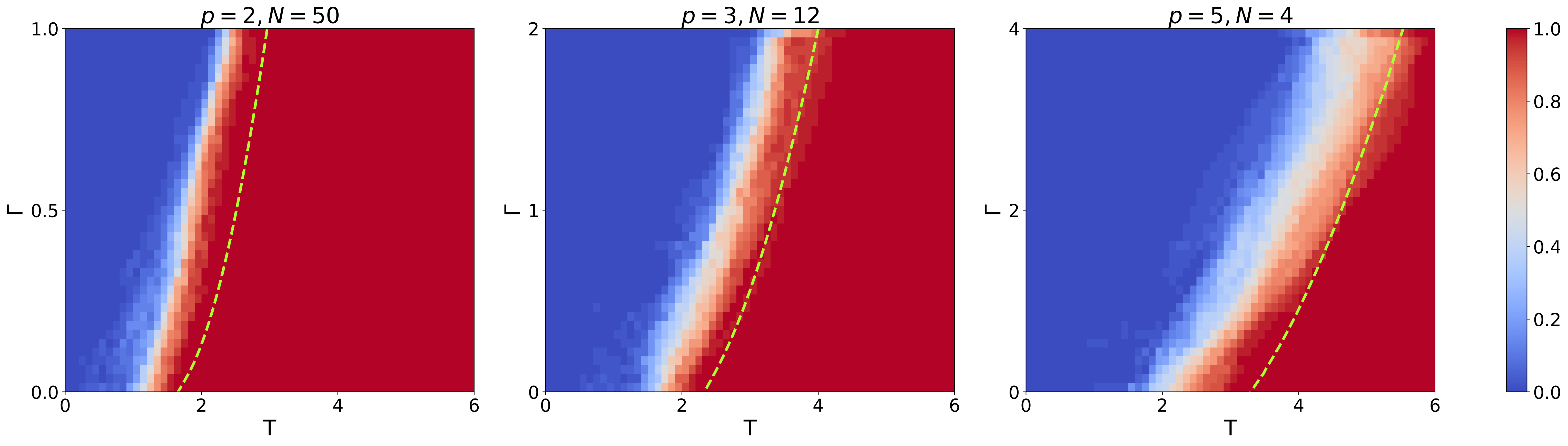}
    \caption{ Heat maps showing the proportion of 40 independent payoff matrices for which all trajectories of Q-Learning converges to a unique fixed point, for varying parameter values. Dark red corresponds to all initial conditions converging to a unique fixed point, while blue indicates there are multiple equilibria. The yellow dashed line is computed from the  generating functional method. It represents the stability boundary, which separates the two regimes in the large action size limit $N \to \infty$.}
    \label{fig:up_fp_dist}
\end{figure*}
~\\
\noindent \textbf{Discussion.}
We are interested in the coordination setting $\Gamma > 0$. 
To generate comparison for games of varying number of players, $p$, we rescale the correlation term as $\hat{\Gamma} = \Gamma / (p-1)$. We will be looking at multi-player games for $\hat{\Gamma} \in (0, 1)$.

Figure \ref{fig:stab_curve} displays the stability curves for varying values of $p$ obtained by solving \eqref{contra2}.  The curve represents the boundary, which separate the multiple fixed point regime from the unique fixed point regime for varying $\hat{\Gamma}$ and $T$.   The key result is as $p$ and $\hat{\Gamma}$ increases, so does the critical exploration rate. What does the boundary look like as $p$ gets larger?  Upon a rescaling the exploration rate by $1/ \sqrt{e(p-1)}$ in the stability plots, Figure \ref{fig:scaled_stab_curve} suggests, as $p$ increases, a direct linear relationship between the critical exploration rate $T_{\text{crit}}$ and  how correlated the game is, $\hat{\Gamma}$, emerges given by \footnote{This linear relationship between critical parameters on the stability boundary is similar to what is observed in DMFT analysis of random replicator predator-prey models in \citep{galla2018dynamically, galla2024generatingfunctionalanalysisrandomlotkavolterra}, subject to a transformation of $\Gamma$ and $T$.}:
\begin{align} \label{estimate_T_crit}
    T_{crit} = (\hat{\Gamma} + 1) \sqrt{e(p-1)} .
\end{align}
In \cite{sanders2018prevalence}, it is shown that in the large-$p$ limit of uncorrelated games ($\hat{\Gamma}= 0$) and $p$-player zero sum games ($\hat{\Gamma}= 1/(p-1)$) the critical exploration rate is given by $T_{\text{crit}} = \sqrt{e(p-1)}$.
This, in combination with \eqref{estimate_T_crit}, suggests the following statement:

\begin{Observation}The critical exploration rate which guarantees the convergence of Q-Learning to a unique fixed point in pure coordination (identical-payoffs) games is \emph{twice} that of 
$p$-player, zero-sum games in the large $p$-limit.
\end{Observation}
\noindent \textbf{Comparison between theory and numerical results.}
We compare how our theoretical results fare against numerical experiments of finite-sized, coordination games. We  used the default SciPy Runga-Kutta 4(5) solver \citep{2020SciPy-NMeth} with max stepsize set to $0.5$ to be approximate continuous Q-Learning \eqref{eq:cql} as closely as possible.  A point is classified as fixed when the derivative of 
\eqref{eq:cql} drops below $|10^{-8}|$.

To determine if a given game converges to a unique fixed point, 100 random initial strategies are drawn and simulated for up to 5000 time units, or until it reaches a fixed point.  If all 100 final points, are within a relative distance of $0.01$ of each other, we assume there is 
a unique fixed point.

For a $p$-player, $N$-action game, we have $p \times N^p$ payoff elements. Selecting $p= 2 , 3 ,5$ and respectively $N= 50, 12 , 4$, yields games with approximately 5000 payoff elements each. For each of the three $(p,N)$-pairs, we perform a parameter search or `mesh-grid' evaluation for $\Gamma \geq 0$ and $T \in (0,6)$. For each $\Gamma$ and $T$, 40 independent games are generated, and we record the proportion of games for which Q-Learning converges to a unique fixed point. 
Figure \ref{fig:up_fp_dist} displays a heat map, displaying the likelihood Q-Learning convergences to a unique fixed point,
given the chosen parameters. 
This is plotted in contrast to the theoretical stability boundary in yellow. We refer to \citep{sanders2018prevalence} for a similar comparison between the theoretical and numerical results, for $\Gamma < 0$.

We can identify a correspondence between the theoretical curve and the simulation results, which validate the generating functional approach. The simulation plots for $p=3$ $N=12$, $p=5$ $N= 4$ has greater variation between sample rounds than $p=2$ $N=50$. 
We suspect increasing the number of actions $N$ should reduce the variation in the dynamics between large games drawn from the same parameters.   
Similar to previous work on competitive games \citep{sanders2018prevalence}, the theoretical critical exploration rate, $T_{\text{crit}}$, appears to be an overestimate, especially near $\Gamma = 0$. We assume (as in previous work) this is a finite-size effect, which disappears as $N$ increases, as the theoretical prediction is in the limit $N \to \infty$.

\section{Conclusion}
Throughout this paper, we have studied the dynamical behaviour of Q-Learning over large, multi-player coordination games, generated from a multivariate Gaussian. This work builds on the model and analysis introduced in \citep{sanders2018prevalence} used to study competitive games, to cover the coordination setting.

Q-Learning in large coordination games exhibits a phenomenon that we call asymptotic extinction, where a non-zero fraction of strategies are played with zero probability in the large action size limit $N \to \infty$. Asymptotic extinction is most noticeable at lower exploration rates $T$, but also occurs at high values of $T$. Taking this effect into account, a critical exploration rate $T_{\text{crit}}$ can be identified above which a unique equilibrium exists, and where all trajectories of Q-Learning from all initial points end up in the same equilibrium. 

The problem of choosing the `optimum' exploration rate remains confounding question. Picking the rate $T_{\text{crit}}$, ensures convergence to a unique distribution, avoiding ending up in worst-case scenarios of converging to bad equilibria. 
$T_{\text{crit}}$ could be taken as a reasonable choice of exploration rate because it is the smallest one where such a unique fixed point is guaranteed, but we emphasize that there are further intriguing questions around the topic of determining the ideal $T$, and there are potentially reasonable alternative choices for $T$. One can consider the problem of finding the exploration rate that maximises any arbitrary objective function (such as maximising total utility, or maximising the minimum utility among the players ). This gives rise to a number of interesting questions to consider for future research.

\section*{Acknowledgments}
The authors would like to thank Aamal Hussain and Edward Plumb for the useful discussions throughout the project. This work was partially supported by the UKRI Trustworthy Autonomous Systems Hub (EP/V00784X/1).  Partial financial support has been received from the Agencia Estatal de Investigaci\'on and Fondo Europeo de Desarrollo Regional (FEDER, UE) under project APASOS (PID2021-122256NB-C21/PID2021-122256NB-C22), and the Maria de Maeztu project CEX2021-001164-M, funded by MCIN/AEI/10.13039/501100011033. Bart de Keijzer was partially supported by EPSRC grant EP/X021696/1.
\bibliography{aaai25}

\onecolumn
\section*{Appendix for Asymptotic Extinction in Large Coordination Games }
\renewcommand\thefigure{A.\arabic{figure}}  
\renewcommand{\theequation}{A.\arabic{equation}}
\setcounter{equation}{0} 
This Appendix provides additional details related to the work presented in the main paper, alongside supporting figures.  The content of the Appendix is ordered as follows:
\begin{enumerate} 
    \item  Motivation for the choice of Gaussian distribution.
    \item  Derivation of the  continuous Q-Learning equation.
    \item  Rescaling of Variables
    \item  Derivation of the effective dynamics.
    \item  Solving fixed points relations for correlated games. $\Gamma \neq 0$
    \item Solving fixed point relation of uncorrelated games $\Gamma = 0$ 
    \item Derivation of the stability condition.
    \item  Further plots of selected simulations.
\end{enumerate}
\subsection*{Motivation for the choice of Gaussian distribution}
 The choice of a Gaussian distribution to describe the distribution of payoff values can be motivated by a maximum-entropy argument. We wish to characterise the outcome of learning in terms of a small number of summary statistics (in our case, the first and second moments of the distribution of payoff matrix elements). Following principles of information theory, one then maximises uncertainty (Shannon entropy) subject to these moments leading to a multivariate Gaussian.\\

We expect our theory to apply for many non-Gaussian distributions. The `universality principle' in random matrix theory \cite{tao2011random} states that the spectra of large non-Gaussian random matrices are identical to those of a Gaussian ensemble with the same first and second moments if higher-order moments fall off sufficiently quickly with the matrix size. The key steps of averaging over the randomness in our calculation is very similar to those in random matrix theory. Hence, we would expect similar universality properties in our model.
\subsection*{Derivation of the continuous Q-Learning equation}
The Q-Learning equations typically given by the following two steps \citep{watkins1992q}: i) updating of the Q-values  
\begin{align}\label{eq:qqlearning}
    Q_a^i(t+1) 
    &= 
    \underbrace{(1- \alpha) Q_a^i(t)}_{\text{discounted previous Q-value}}
    + 
    \underbrace{R(a, \mathbf{x}^{-i}(t))^i}_{\text{current reward}}
\end{align}
ii) applying a softmax to obtain strategies from the Q-values:
\begin{align}\label{eq:ssoftmax}
    x_a^i(t) & =\frac{\exp \left[\beta Q_a^i(t)\right]}{\sum_{b \in \mathcal{A}} \exp \left[\beta Q_b^i(t)\right]} 
\end{align}
where
the discount rate is then given by $(1-\alpha)$ and  the softmax distribution parameterised by $\beta>0$
Combining \eqref{eq:qqlearning} and \eqref{eq:ssoftmax}, we obtain the following recursive relation for $x$:
\begin{align}\label{eq:dis_q}
    x_a^i(t+1) & =\frac{1}{\hat{\rho} ^i (t)} x_a^i(t)^{(1-\alpha)} \left[\exp(\beta R(a, \mathbf{x}^{-i}(t))^i)\right]
\end{align}
where $\hat{\rho}^i (t):= \sum_{k \in \mathcal{A}} x^i_k(t)^{1-\alpha} \left[\exp(\beta R(a, \mathbf{x}^{-i}(t))^i)\right]$ is a normalisation parameter, which ensures strategies stay within the simplex.  Typically the step sizes between each update in reinforcement learning is small, this motivates finding the continuous Q-Learning equations, an ODE which represents the small stepsize limit. Dividing both sides sides by $x_a^i(t)$ and taking the logarithm we obtain the following relation:
\begin{align}\label{eq:dis_q1}
    \ln \frac{x_a^i(t+1)}{x_a^i(t)} & =  \beta R(a, \mathbf{x}^{-i}(t))^i) -\alpha \ln x_a^i(t) - \ln \hat{\rho}^i (t)
\end{align}
With an abuse of notation, in the small step size limit, the update between each discrete time step is minimal (i.e. we have $t+1 \approx t$).  This allows us to make the following approximation:
\begin{align} \label{approx}
\frac{\ln x_a^i(t+1) - \ln x_a^i(t)}{(t+1) -t } \approx \frac{d}{dt} \ln x_a^i(t) = \frac{\dot{x}_a^i(t)}{x_a^i(t)}
\end{align}
As discussed in the main paper, we refer to parameter $T:= \alpha / \beta $ as the \emph{exploration rate}.  Taking $\alpha, \beta \to 0$, but keeping $T$ constant is equivalent to taking smaller step size in each update until we reach the continuous limit. Substituting the approximation \eqref{approx} into \eqref{eq:dis_q1} and sending  $\alpha, \beta \to 0$, we obtain the continuous Q-Learning equations \cite{sato2003coupled, tuyls:qlearning}:
\begin{align}\label{cql}
    \frac{\dot{x}_a^i(t)}{x_a^i(t)}=R(a, \mathbf{x}^{-i}(t))^i - T \ln x_{a}^i(t) - \rho^i(t) , 
\end{align}
where  $\rho^i:= R(\vec{x}^i(t), \mathbf{x}^{-i}(t))^i - T \langle \vec{x}^i(t), \ln \vec{x}^i(t) \rangle$ is the new normalisation parameter and  $\langle\cdot, \cdot \rangle$ denotes the inner product.
\subsection*{Rescaling of Variables}
The equation \eqref{cql} holds for any value of $N$.  In this segment. we provide an overview of the rescaling of variables introduced in \citep{sanders2018prevalence}.  To understand why a rescaling has to be done, we will first write down how the various terms in our equation scale w.r.t with $N$:
\subsubsection{Rescaling of $T$}
\begin{itemize}
    \item[--] We expect the typical action to be played probability $1/N$,  $x_a^i(t) \sim O(1/N)$.
    \item[--] The typical entry in the payoff matrix is not dependent on $N$, $\Pi(\vec{a})_i\sim O(1)$.
\end{itemize}
Let us consider player $i$'s perspective. They wish to update their strategy using the Q-Learning algorithm.  This requires calculating the expected reward of each pure action.
Holding player $i$'s strategy fixed as a pure action, the typical reward contribution from an action profile $\vec{a}$ (or each entry in the payoff matrix) given by $\Pi(\vec{a})_i \prod_{j \in \mathcal{P}, i \neq j} x^j_{\vec{a}_j}$ scales at a rate of $ O(1/N^{p-1})$.

From player $i$'s point of view, upon fixing their action, the remaining players could play one among $N^{p-1}$ action profiles.  Since the reward associated with each action profile is independent, the Q-value ,$Q_a^i(t)$, and the expected reward for a given action given by:  
$R(a, \mathbf{x}^{-i}(t))_i = \sum_{\vec{a} \in \mathcal{A}^p , i \neq j}
\Pi(\vec{a})_i \prod_{i \in \mathcal{P}, i \neq j} x^i_{a_i}$ has a standard deviation of $O \left(\sqrt{1/ N^{(p-1)}}\right)$. 
This means that as $N$ and $p$ increases, the difference in rewards across different actions become less distinguishable, going to $0$ in the limit.  For exploitation to remain meaningful, the players have to adjust their exploration rate accordingly.  Since $Q_a^i(t)$ has a standard deviation of $O \left(\sqrt{1/ N^{(p-1)}}\right)$, $\beta$ in the soft-max $\eqref{eq:softmax}$ should scale with a factor $\sqrt{N^{(p-1)}}$. This yields us the change of variables for the exploration rate in the main paper given by:
$$T = \tilde{T}/\sqrt{N^{(p-1)}}$$
\subsubsection{Rescaling of $x$}
We note that this equivalent to keeping $T$ fixed and rescaling $x$ by a factor of $N$ and the matrix elements  by $\sqrt{1/ N^{(p-1)}}$.  We introduce $\tilde{x}_a^i = N x_a^i  \sim O(1)$ and $\tilde{\Pi}(\vec{a})_{i} = \Pi(\vec{a})_{i} /\sqrt{N^{(p-1)}} \sim O(\sqrt{1/ N^{(p-1)}})$ such that the payoff correlations take the following form:
\begin{align*}
    \mathbb{E}\left[
    \tilde{\Pi}(\vec{a})_{i} \cdot 
    \tilde{\Pi}(\vec{b})_{j}
    \right]
    = 
\begin{cases}
1/N^{(p-1)}, & \text{ if } \vec{a} = \vec{b}, i = j , \\
(p-1)^{-1} \Gamma /N^{(p-1)}, & \text{ if }  \vec{a} = \vec{b}, i \neq j \\
0, & \text{ if } \vec{a} \neq \vec{b}
\end{cases}
\end{align*}
Now the `typical contribution' for each action profile scales at a rate of $\Pi(\vec{a})_i \prod_{j \in \mathcal{P}, i \neq j} x^j_{\vec{a}_j} \sim O \left(\sqrt{1/ N^{(p-1)}}\right)$ and thus, the typical reward and Q values scales with $O(1)$, meaning we do not have to change $T$ w.r.t $N$ under this particular rescaling.  In the derivation of the effective dynamics, we will drop the tildes and use this second set of rescaling on $x$ and $\Pi(\vec{a})$.
\subsection*{Derivation of the effective dynamics}
In this section, we will present the derivation of the effective dynamics presented in the Supplementary of  \citep{sanders2018prevalence}. For a step-by-step approach on this method on replicator models, refer to \citep{galla2024generatingfunctionalanalysisrandomlotkavolterra}.
We will break up this segment into the following parts:
\begin{itemize}
    \item[--] Introduce the generating function induced by Q-Learning.
    \item[--] Identify the effective dynamics of ensembles of games under Q-Learning.
\end{itemize}
\newcommand{\at}[2][]{#1|_{#2}}

\subsubsection{Generating  Functions}
We begin with the Characteristic Function of a univariate random variable $X$, which is given by:
\begin{align}\label{cgf}
    Z_X(t) = \mathbb{E} (e^{itX}) =\int f_X(X) e^{itX} dX 
\end{align}
where $f_x(x)$ is the pdf of $X$ and $t \in \mathbb{R}$ is the parameter of the generating function.  It takes Fourier Transform of the probability density function.  It has a some very neat properties namely, the n-th order moment  can be found by taking the derivatives 
\begin{align*}
    \mathbb{E}(X^n) &= i^{(-n)}\frac{d^n}{dt^n}Z_X(t)\at[\big] {t=0}
\end{align*}
For a multivariate time series, $\vec{x}  = (x_t^i , t \in [t_0 , T],  \in[0, 1 \dots n] )$.  The characteristic function takes the following form:
\begin{align} \label{cgf_time_series}
    Z_\vec{x} (\psi) =& \mathbb{E} \left[\exp{ \left( i \sum_i \int{ \psi^i(t) x^i_t dt }\right) } \right]=\int D \vec{x} P(\vec{x}) \exp{ \left( i \sum_i \int{ \psi^i(t) x^i_t dt }\right) }  
\end{align}
where $P(\vec{x} )$ is the probability measure over all possible trajectories and $\psi^i(\cdot):[t_0, T) \to \mathbb{R}$. $\psi^i(t)$ replaces $t$ in \eqref{cgf} to become the parameter associated with $x^i(t)$.  The following moment property describes time correlations across different variables:
\begin{align}\label{cgf_moments}
    \mathbb{E} \left[x^i(t) x^j(t') \right] = - 
    \frac{\partial^2}{\partial {\psi^i(t)} \partial {\psi^j(t')}}
    Z_\vec{x} (\psi)\at[\big] {\psi^i(t) =\psi^j(t') =0}
\end{align}
Moving forward, we want to keep the moment property \eqref{cgf_moments} in mind, as we derive a generating function associated with Q-Learning.  Consider the following modified continuous Q-Learning equations:    
\begin{align}
    \frac{\dot{ x}_a^i(t)}{x_a^i(t)}= R(a, \mathbf{x}^{-i}(t))^i - T \ln x_{a}^i(t) - \rho^i(t)  + h_a^i(t)
    \label{eq:mod_cql} 
\end{align}
This identical the continuous Q-Learning equations \eqref{cql}, except we have included an arbitrary function $h_a^i(t)$, this enables us to generate a response function (and which we will be set zero).  The generating function for Q-Learning is then given to take the following form:
\begin{align}
    Z_{\vec{x}}(\psi )= \int D \vec{x} \;  \delta(\text{eq. of motion})
            \exp{\left( i \sum_{a,i} \int dt 
             \psi^i_a(t) x^i_a(t) 
           \right)} \notag\\
\end{align}
This takes a similar form to \eqref{cgf_time_series}. 
 Here $P(\vec{x})= \delta(\text{eq. of motion})$ describes the probability measure induced by the possible Q-Learning trajectories.  Expanding out the $\delta$-functions with a Fourier Transform we have the generating function induced by Q-Learning:
\begin{align} \label{gen_fun_q_learning}
     Z_{\vec{x}}(\psi ) =& \int D[\vec{x}, \hat{\vec{x}}] 
            \exp{\left( i \sum_{a,i} \int dt 
           \left[
           \hat{x}^i_a(t) 
           \left(
           \frac{\dot{ x}_a^i(t)}{x_a^i(t)} -\left( R(a, \mathbf{x}^{-i}(t))^i - T \ln x_{a}^i(t) - \rho^i(t)  + h_a^i(t)
           \right)
           \right)
           \right] + \psi^i_a(t) x^i_a(t) 
           \right)}
\end{align}
Similar to \eqref{cgf_moments}, \eqref{gen_fun_q_learning} admits the following time-correlation and response function relations:
\begin{align}
    \mathbb{E} \left[x_a^i(t) x_b^j(t') \right] &= 
    - 
    \frac{\partial^2}{\partial {\psi^i_a(t)} \partial {\psi^j_b(t')}}
    Z_\vec{x} (\psi)\at[\big] {\vec{\psi}= \vec{h} =0} \notag \\
    \frac{\partial } {\partial h^j_b(t^\prime)}\mathbb{E} \left[ x_a^i(t) \right] &= 
    - 
    \frac{\partial^2}{\partial {\psi^i_a(t)} \partial {h^j_b(t^\prime)}}
    Z_\vec{x} (\psi)\at[\big] {\vec{\psi}= \vec{h} =0} \label{moment_props}
\end{align}

\subsubsection{Identifying the effective dynamics
}
The randomness associated with  \eqref{gen_fun_q_learning} comes from the entries of the payoff matrix.  Separating this from the other elements, we obtain:
\begin{align} \label{gen_fun_separated}
     Z_{\vec{x}}(\psi ) =& \int D[\vec{x}, \hat{\vec{x}}] 
            \exp{\left( i \sum_{a,i} \int dt 
           \left[
           \hat{x}^i_a(t) 
           \left(
           \frac{\dot{ x}_a^i(t)}{x_a^i(t)} -\left( - T \ln x_{a}^i(t) - \rho^i(t)  + h_a^i(t)
           \right)
           \right)
           \right] + \psi^i_a(t) x^i_a(t) 
           \right)} \notag\\
           & \times \underbrace{\exp{\left( i \sum_{a,i} \int dt \hat{x}^i_a(t)  R(a, \mathbf{x}^{-i}(t))^i \right)}}_{\text{randomness comes from here}}
\end{align}
We will now conduct an averaging over all possible payoff matrices. Let us evaluate the expectation from the final row, which we denote as $Z_\Pi$. To do this, we have to integrate w.r.t to the probability measure of the multivariate Gaussian corresponding to the payoff entries (which we denote as $P (\vec{\Pi(a)})$.
\begin{align}
    \overline{Z_\Pi} = \mathbb{E} \left[\exp{\left( i \sum_{a,i} \int dt \hat{x}^i_a(t)  R(a, \mathbf{x}^{-i}(t))^i \right)} \right] = \int \exp{\left( i \sum_{a,i} \int dt \hat{x}^i_a(t)  R(a, \mathbf{x}^{-i}(t))^i \right)} P (\vec{\Pi(a)}) D[\vec{\Pi(a)}]
\end{align}
Noting the characteristic function of a multivariate Gaussian is given by:
\begin{align}
    \mathbb{E} \left[ \exp{\left(itX\right)} \right] = 
\exp{\left(i \mathbf{\mu ^T t} - \frac{1}{2}  \mathbf{t^T \Sigma t} \right)} \; , \; X \sim \mathcal{N}(\mu,\,\Sigma^{2}) 
\end{align}
From the rescaling in the previous segment, the rescaled payoff matrix correlations are as follows:
\begin{align*}
    \mathbb{E}\left[
    \Pi(\vec{a})_{i} \cdot 
    \Pi(\vec{b})_{j}
    \right]
    = 
\begin{cases}
1/N^{(p-1)}, & \text{ if } \vec{a} = \vec{b}, i = j , \\
(p-1)^{-1} \Gamma /N^{(p-1)}, & \text{ if }  \vec{a} = \vec{b}, i \neq j \\
0, & \text{ if } \vec{a} \neq \vec{b}
\end{cases}
\end{align*}
$\overline{Z_\Pi}$ can be rewritten as: 
\begin{align} \label{z_pi}
    \overline{Z_\Pi}= \exp \left( -\frac{N}{2} \int d t d t^{\prime} \sum_{i}
    \left(
    L^i(t,t^{\prime}) \prod_{i \neq j} C^j(t,t^{\prime})
    + \Gamma \sum_{i \neq j}K^i(t,t^{\prime}) K^j(t,t^{\prime})
    \prod_{k \notin (i,j)} C^k(t,t^{\prime})
    \right)
    \right)
\end{align} 
where the following short-hand substitutions are introduced:  
\begin{align*}
C^i\left(t, t^{\prime}\right)
&=
\frac{1}{N} \sum_{a} x^i_{a}(t) x^i_{a}\left(t^{\prime}\right) \\
K^{i}\left(t, t^{\prime}\right)
&=
\frac{1}{N} \sum_{a} x^i_{a}(t) \widehat{x}^i_{a}\left(t^{\prime}\right)\\
L^{i}\left(t, t^{\prime}\right)
&=
\frac{1}{N} \sum_{a} \widehat{x}^i_{a}(t) \widehat{x}^i_{a}\left(t^{\prime}\right)
\end{align*}
We will be substituting \eqref{z_pi} with the bottom row of \eqref{gen_fun_separated}, to complete averaging step: 
\begin{align} \label{gen_fun_separated2}
     \overline{Z_{\vec{x}}(\psi )} =& \int D[\vec{x}, \hat{\vec{x}}] 
            \exp{\left( i \sum_{a,i} \int dt 
           \left[
           \hat{x}^i_a(t) 
           \left(
           \frac{\dot{ x}_a^i(t)}{x_a^i(t)} -\left( - T \ln x_{a}^i(t) - \rho^i(t)  + h_a^i(t)
           \right)
           \right)
           \right] + \psi^i_a(t) x^i_a(t) 
           \right)} \times \overline{Z_\Pi}
\end{align}
We re-express $Z_{\vec{x}}(\psi )$ as an integration over the short-hand substitutions terms and their respective conjugates.  We seek the following form for $Z_{\vec{x}}(\psi )$; turning equation \eqref{gen_fun_q_learning} to the form of \eqref{conjugate form}.
\begin{align}
     Z_{\vec{x}}(\psi) &= \int D[\vec{x}, \hat{\vec{x}}] \times \dots \tag{A.12} \\
     & \to \int D[\vec{x}, \hat{\vec{x}} ] \times D\left[C, L, K, \widehat{C}, \widehat{L}, \widehat{K}\right] \times \dots \label{conjugate form}
\end{align}
We note:
\begin{align}
    1 &= \int D[C^i] P(C^i) \delta(0) \label{step1}\\
    &= \int D[C^i] \prod_{t,t'} \delta 
    \left( \underbrace{C^i\left(t, t^{\prime}\right) - \frac{1}{N} \sum_{a} x^i_{a}(t) x^i_{a}(t')}_{= 0} \right)\label{step2}\\
    &= \int D[\widehat{C}^i, C^i] \exp\left( iN \int dt dt'\left( \widehat{C}^i\left(t, t^{\prime}\right) C^i\left(t, t^{\prime}\right) - \frac{\widehat{C^i}\left(t, t^{\prime}\right)}{N} \sum_{a} x^{i}_a(t) x^{i}_a(t') \right) \right) \label{step3}
\end{align}
where:
\begin{itemize}
    \item[--] \eqref{step1} states that the integral of the pdf of $C_i = 1$.
    \item[--] \eqref{step2} rewrites \eqref{step1}
    \item[--] \eqref{step3} introduces the conjugate integration variable.
\end{itemize}
Repeating the expansions \eqref{step1} - \eqref{step3} for the other short-hand terms and multiplying onto \eqref{gen_fun_q_learning} we can express the generating functional in the conjugate form as described in \eqref{conjugate form}:
\begin{align} \label{avg_conjugate}
    \overline{Z_{\vec{x}}(\psi )} &= \int D\left[C, L, K, \widehat{C}, \widehat{L}, \widehat{K}\right] \exp(F)
\end{align}
where:
$$F= N ( \Psi + \Phi + \Omega)$$
denotes the terms in the exponent given by:
\begin{align} \label{psipsi}
    \Psi &= i \sum_i \int dt  dt^{\prime} 
    \left(
    \widehat{C}^i\left(t, t^{\prime}\right) C^i \left(t, t^{\prime}\right)+
    \widehat{K}^i\left(t, t^{\prime}\right) K^i\left(t, t^{\prime}\right)+
    \widehat{L}^i\left(t, t^{\prime}\right) L^i\left(t, t^{\prime}\right)  \right)
\end{align}
comes from the introduction of the conjugate variables in \eqref{step3}.
\begin{align} \label{z_pi_mod}
    \Phi=   -\frac{1}{2}\sum_{i} \int d t d t^{\prime} 
    \left(
    L^i(t,t^{\prime}) \prod_{i \neq j} C^j(t,t^{\prime})
    + \Gamma \sum_{i \neq j}K^i(t,t^{\prime}) K^j(t,t^{\prime}
    )\prod_{k \notin (i,j)} C^k(t,t^{\prime})
    \right)
 \end{align} 
comes from $\overline{Z_\Pi}$
\begin{align} \label{omegaomega}
    \Omega =& N^{-1} \sum_{i,a} \ln \left[\int D\left[x_{a}^i, \widehat{x}_{a}^i\right] p_{a, t_0}^{(i)}\left(x^i_{a}(t_0)\right) \exp \left(i \int d t \psi^i_a(t) x^i_a(t) \right)\right. \notag\\
    & \times \exp{ \left(i \sum_{a,i} \int dt 
           \left[
           \hat{x}^i_a(t) 
           \left(
           \frac{\dot{ x}_a^i(t)}{x_a^i(t)} -\left(   - T \ln x_{a}^i(t) - \rho^i(t)  + h_a^i(t)
           \right)
           \right)
           \right]\right)} \notag\\
    &  \left.\times \exp \left(-i \int d t d t^{\prime}\left[
    \widehat{C}^i\left(t, t^{\prime}\right) x_{a}^i(t) x_{a}^i\left(t^{\prime}\right)
    +
    \widehat{K}^i\left(t, t^{\prime}\right) x_{a}^i(t)  \widehat{x}^i_{a}(t^{\prime}) 
    +
    \widehat{L}^i\left(t, t^{\prime}\right) 
    \widehat{x}_{a}^i(t) \widehat{x}^i_{a}(t^{\prime}) 
    \right]\right) \right]
\end{align}
 contains the remaining terms notably, $D[\vec{x}, \hat{\vec{x}}]$. We have introduced $p_{a, t_0}^{(i)} (\cdot)$ to represent the initial distribution at $t_0$. \eqref{avg_conjugate} takes the form: $I = \int dx \exp(f(x))$, in the $N \to \infty$ limit the saddle point approximation holds (See \citep{asc_peng} under Laplace's method) given by:
 \begin{align} \label{saddle
 }    I\approx \exp(f(\hat{x})) \int dx \exp{\left(\frac{1}{2} (x - \hat{x})^2 f''(\hat{x}) \right)}
\end{align}
where $\hat{x}$ is the value of $x$ which maximises $f$ i.e.$ f(\hat{x}) = f_{max}$.  This approximation is found by considering the second-order Taylor expansion around $\hat{x}$ and most of the weight of the integral is near the maxima.  Finding the maxima corresponding to the terms in the exponent for \eqref{avg_conjugate}, this involves taking the partial derivatives of $F$ w.r.t $\left[C, L, K, \widehat{C}, \widehat{L}, \widehat{K}\right]$.
\begin{align*}
    0= \frac{\partial}{\partial C^i} F = \frac{\partial}{\partial K^i}F = \frac{\partial}{\partial L^i} F= \dots
\end{align*}
This gives the following relations from taking derivatives of $C, K, L$:
\begin{align}
    i \widehat{C}^i(t, t^{\prime}) &= \frac{1}{2} \sum_{j \neq i}
    \left(
    L^j(t,t^{\prime}) \prod_{k \notin (i,j)} C^k(t,t^{\prime})
    + 
    \Gamma \sum_{k \notin (i,j)}K^j(t,t^{\prime}) K^k(t,t^{\prime}
    )\prod_{l \notin (i,j,k)} C^l(t,t^{\prime})
    \right) \notag\\
    i \widehat{K}^i(t, t^{\prime}) &= 
    \Gamma \sum_{j\neq i} K^j(t,t^{\prime}) \prod_{k \notin (i,j)} C^k(t,t^{\prime})
    \notag\\
    i \widehat{L}^i(t, t^{\prime}) &= 
    \frac{1}{2} \prod_{j \neq i} C^j(t,t^{\prime}) \label{relation_non_conj}
\end{align}
While taking the derivate w.r.t the conjugate variables $\widehat{C}, \widehat{L}, \widehat{K}$ yields:
\begin{align}
    C^i\left(t, t^{\prime}\right) & 
    =\lim _{N \rightarrow \infty} N^{-1} \sum_{a}\left\langle x_{a}^i(t) x_{a}^i\left(t^{\prime}\right)\right\rangle_{\Omega} \notag\\
    K^i\left(t, t^{\prime}\right) &
    =\lim _{N \rightarrow \infty} N^{-1} \sum_{a}\left\langle x_{a}^i(t) \widehat{x}_{a}^i\left(t^{\prime}\right)\right\rangle_{\Omega} \notag\\
    L^i\left(t, t^{\prime}\right) & 
    =\lim _{N \rightarrow \infty} N^{-1} \sum_{a}\left\langle \widehat{x}_{a}^i(t) \widehat{x}_{a}^i\left(t^{\prime}\right)\right\rangle_{\Omega} \label{relation_conj}
\end{align}
where the average $\langle\ldots\rangle_{\Omega}$ is the average over the probability distribution defined by $\Omega$ \eqref{omegaomega}.  
We note from the properties of the generating functional \eqref{moment_props}, we have the following:
 \begin{align*}
& C^i\left(t, t^{\prime}\right)=
-\left.\lim _{N \rightarrow \infty} 
N^{-1} \sum_{a} \frac{\partial^{2} Z_{\vec{x}}(\psi)}
{\partial \psi_{a}^i(t) \partial\psi_{a}^i\left(t^{\prime}\right)}\right|_{\boldsymbol{\psi}=\mathbf{h}=0}, \\
& K^i\left(t, t^{\prime}\right)=
-\left.\lim _{N \rightarrow \infty} N^{-1} 
\sum_{a} \frac{\partial^{2} Z_{\vec{x}}(\psi)}{\partial \psi^i_{a}(t) \partial h_a^i\left(t^{\prime}\right)}\right|_{\boldsymbol{\psi}=\mathbf{h}=0}, \\
& L^i\left(t, t^{\prime}\right)=
-\left.\lim _{N \rightarrow \infty} N^{-1}
\sum_{a} \frac{\partial^{2} Z_{\vec{x}}(\psi)}{\partial h_a^i\left(t\right) \partial h_a^{ i}\left(t^{\prime}\right)}\right|_{\boldsymbol{\psi}=\mathbf{h}=0}
\end{align*}
We note the following: $Z_{\vec{x}} \left[\psi = 0 \right] = 1, \; \forall h $ due to normalisation. Thus $L^i\left(t, t^{\prime}\right) = 0, \; \forall i ,t , t^\prime$.  Due to causality, we have $ K^i\left(t, t^{\prime}\right) = 0, \; \forall t^\prime < t$.  Making the appropriate substitutions, the $\Psi + \Phi$ terms disappear and we are left with $F = N \Omega$.  Assuming a the perturbations $h_a^i(t) = h(t)$ are symmetric with respect to each player and action alongside the initial probability distributions $p_{a, t_0}^{(i)} = p_{t_0}$, we drop the dependence of have on $a,i$ to obtain:
\begin{align}
\Omega= & p \ln \left\{\int \mathcal{D}[x, \widehat{x}] p_{t_0}(x(t_0)) \exp \left(\mathrm{i} \int \mathrm{d} t \widehat{x}(t)\left(\frac{\dot{x}(t)}{x(t)}+T \ln x(t)+\rho(t)-h(t)\right)\right)\right. \notag \\
& \left.\times \exp \left[-\int \mathrm{d} t \int \mathrm{d} t^{\prime}\left(\Gamma(p-1) K\left(t, t^{\prime}\right) C\left(t, t^{\prime}\right)^{p-2} x(t) \widehat{x}\left(t^{\prime}\right)+\frac{1}{2} C\left(t, t^{\prime}\right)^{p-1} \widehat{x}(t) \widehat{x}\left(t^{\prime}\right)\right)\right]\right\} \label{new_omega}
\end{align}
Substituting into \eqref{avg_conjugate} and making the substitution $G\left(t, t^{\prime}\right)=-\mathrm{i} K\left(t, t^{\prime}\right)$, we have the following generating function:
\begin{align}
Z_{\text {eff }}= & \int \mathcal{D}[x, \widehat{x}] p_{t_0}(x(t_0)) \exp \left(\mathrm{i} \int \mathrm{d} t \widehat{x}(t)\left(\frac{\dot{x}(t)}{x(t)}+ T \ln x(t)+\rho(t)-h(t)\right)\right) \notag\\
& \times \exp \left[-\int \mathrm{d} t \int \mathrm{d} t^{\prime}\left(\mathrm{i} \Gamma G\left(t, t^{\prime}\right) C\left(t, t^{\prime}\right)^{p-2} x(t) \widehat{x}\left(t^{\prime}\right)+\frac{1}{2} C\left(t, t^{\prime}\right)^{p-1} \widehat{x}(t) \widehat{x}\left(t^{\prime}\right)\right)\right] \label{eff_dyn_gf}
\end{align}
which corresponds to the generating function of the effective dynamic equation
\begin{align}
\frac{\dot{x}(t)}{x(t)}=\Gamma \int \mathrm{d} t^{\prime} G\left(t, t^{\prime}\right) C\left(t, t^{\prime}\right)^{p-2} x\left(t^{\prime}\right)-T \ln x(t)-\rho(t)+\eta(t)+h(t) \label{eff_dyn2}
\end{align}
where $\eta(t)$ is a coloured (i.e., time-correlated) Gaussian random variable satisfying: 
    $\left\langle \eta(t)\eta(t^{\prime}) \right\rangle_{*} = C\left(t, t^{\prime}\right)^{p-1}$. $C$ and $G$ represents the time correlation term and response function respectively and are given by:
\begin{align*}
C\left(t, t^{\prime}\right)= \left\langle x(t) x(t')\right\rangle_{*} , \;G\left(t, t^{\prime}\right)=\left\langle\frac{\partial x(t)}{\partial h\left(t^{\prime}\right)}\right\rangle_{*} .
\end{align*}
where $\left\langle \dots \right\rangle_{*}$ denotes the expected value over realisations of the effective dynamic \eqref{eff_dyn2}.  Since $h(t)$ is defined to be zero for all time, we can drop $h(t)$ to finally obtain the effective dynamics as  we see in the main paper: 
\begin{align}
\frac{\dot{x}(t)}{x(t)}=\Gamma \int \mathrm{d} t^{\prime} G\left(t, t^{\prime}\right) C\left(t, t^{\prime}\right)^{p-2} x\left(t^{\prime}\right)-T \ln x(t)-\rho(t)+\eta(t) \label{eff_dyn_2}
\end{align}
Lastly, the partial derivative of $G$ w.r.t $\eta(t^\prime)$ and $h(t^\prime)$ are equivialent, we rewrite $G$ as follows:
$$G\left(t, t^{\prime}\right)=\left\langle\frac{\partial x(t)}{\partial \eta\left(t^{\prime}\right)}\right\rangle_{*} $$
\subsection*{Solving Fixed points of correlated Games $\Gamma \neq 0$ }
The fixed point relation \eqref{fixed_point_eq} and the associated self-consistency relation \eqref{self_suff} cannot be solved directly, but rather has to be approximated numerically.  We have used a Newton method to find the fixed points by converting the self-consistency relations to loss functions to minimise (do contact the corresponding author for coded  implementation in Python).  A pseudo-code detailing how we found the fixed point relations for a given coorelation parameter, $\Gamma$,  and exploration rate, $T$, is given as follows: 

\begin{algorithm}[H]
\DontPrintSemicolon
  
  \KwInput{$\Gamma$(Coorelation parameter), $T$ (Exploration rate), $p$ (number of players)}
  \KwOutput{$x: z \to \mathbb{R}$ (fixed point distribution of $x$)}
 
 - Let $\hat{x}(z; K, a , b)$ be parameterised by $K,a, b$, as in \eqref{fixed_gamma_neg}, \eqref{fixed_gamma_pos} from the fixed point relations in the main paper.\\\

 We define a general loss function consisting of 3 smaller functions as follows:
$$\text{Loss}(K,a,b)= (\text{loss}_1(K, a, b))^2 + (\text{loss}_2(K, a, b))^2 + (\text{loss}_3(K, a, b))^2$$
{where the sub-functions are defined to be:}
    {\begin{itemize}
        \item[--]  loss$_1$$(K, a, b)$ = $ \mathbf{q}^{(p-1)/2}  - \int^{\infty}_{-\infty} Dz \;
\frac{\delta \hat{x}(z)}{\delta z}  \chi$
    \item[--] loss$_2$$(K, a, b)$ = $\mathbf{q} - \int^{\infty}_{-\infty} Dz \;\hat{x}(z)^2$
    \item[--] loss$_3$$(K, a, b)$ = $1 - \int^{\infty}_{-\infty} Dz \;\hat{x}(z)$ 
    \end{itemize}}
where the integrals are estimated numerically and $\mathbf{q}$ and $\chi$ are determined from $a, b$. Recall:
\begin{itemize}
    \item[--] $a = \Gamma \mathbf{q}^{(p-2)}$
    \item[--] $b = \mathbf{q}^{(p-1)/2} T^{-1}$
    \item[--] $Dz = \exp(-z^2/ 2)/ \sqrt{2 \pi} dz$
\end{itemize}

Given an initial guess, $K_0, a_0, b_0$, we use a standard Newton method to find $\mathbf{K, a, b}$ which minimise following the general loss function:
$$\mathbf{K, a, b}=\text{minimise}(\text{Loss}(K_0,a_0,b_0))$$
For almost any input of $T, \Gamma$ and $p$ (in the relevant regime)\footnote{For parameter values far from the stability boundary, like $T << T_\text{crit}$ and for $\Gamma$ near 0, there might be a higher associated error. },we can obtain a Loss of almost $0$ ($ < 10^{-16}$) at the minima.  This corresponds to finding parameterisation of $x(z; K,a,b$), which almost exactly solves the fixed point relations \eqref{fixed_point_eq} and the associated self-consistency relation \eqref{self_suff}. 
\\
\textbf{Return} {$\hat{x}(z;\mathbf{K, a, b})$}
\caption{Fixed point computation for $\Gamma \neq 0$.}
\end{algorithm}

\subsection*{Solving Fixed points of Uncorrelated Games $\Gamma = 0$}
In this section, we will solve the fixed points of uncorrelated games.  As discussed in the main paper, fixed points can be internal or on the boundary depending on the exploring rate.  We can observe this behaviour change in numerical simulations:\\

\begin{figure*}[t]
    \centering
    \subfigure{\centering
        \includegraphics[width=0.48\linewidth]{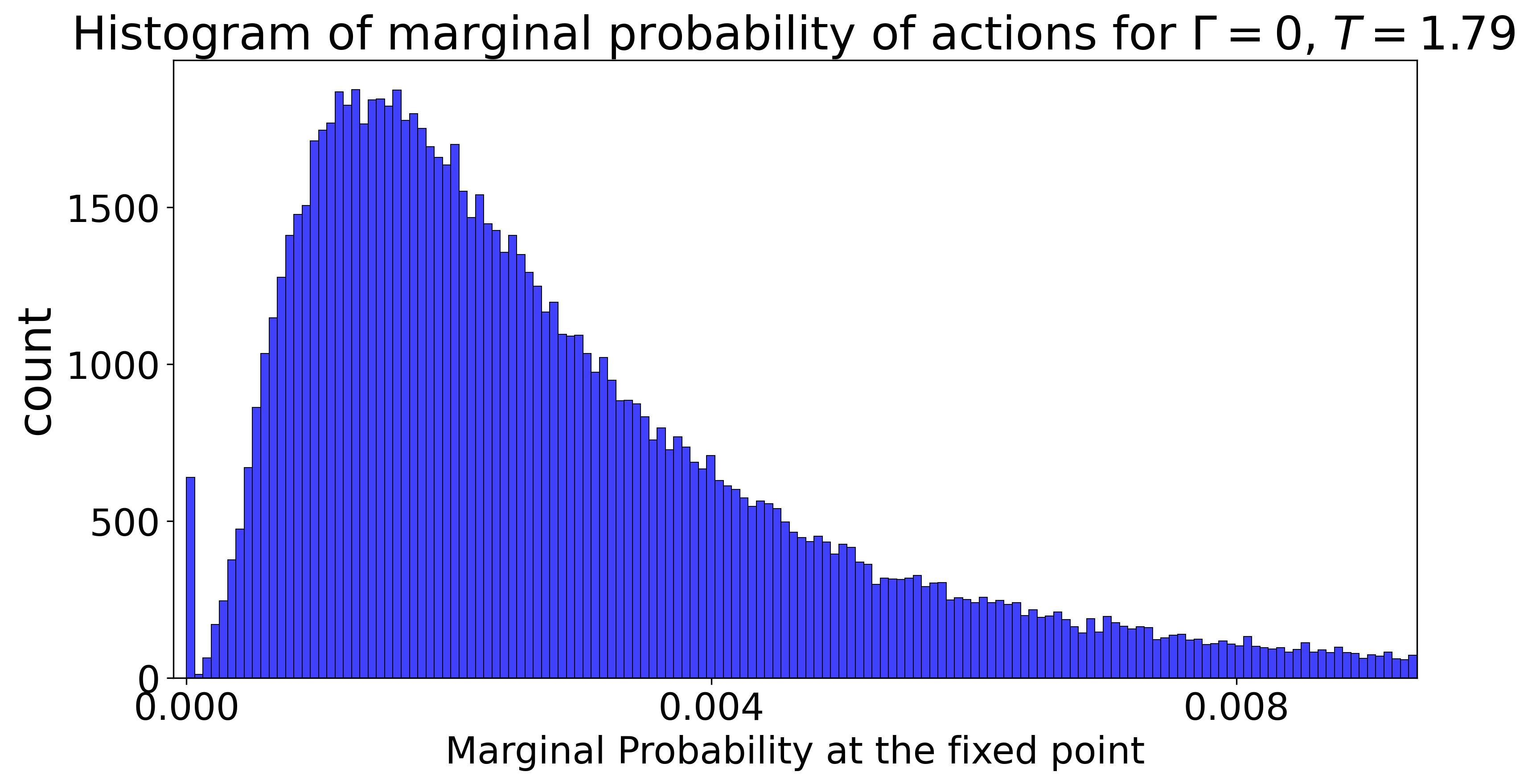}
        \label{fig:boundary0} 
    }
   \subfigure{\centering
        \includegraphics[width=0.48\linewidth]{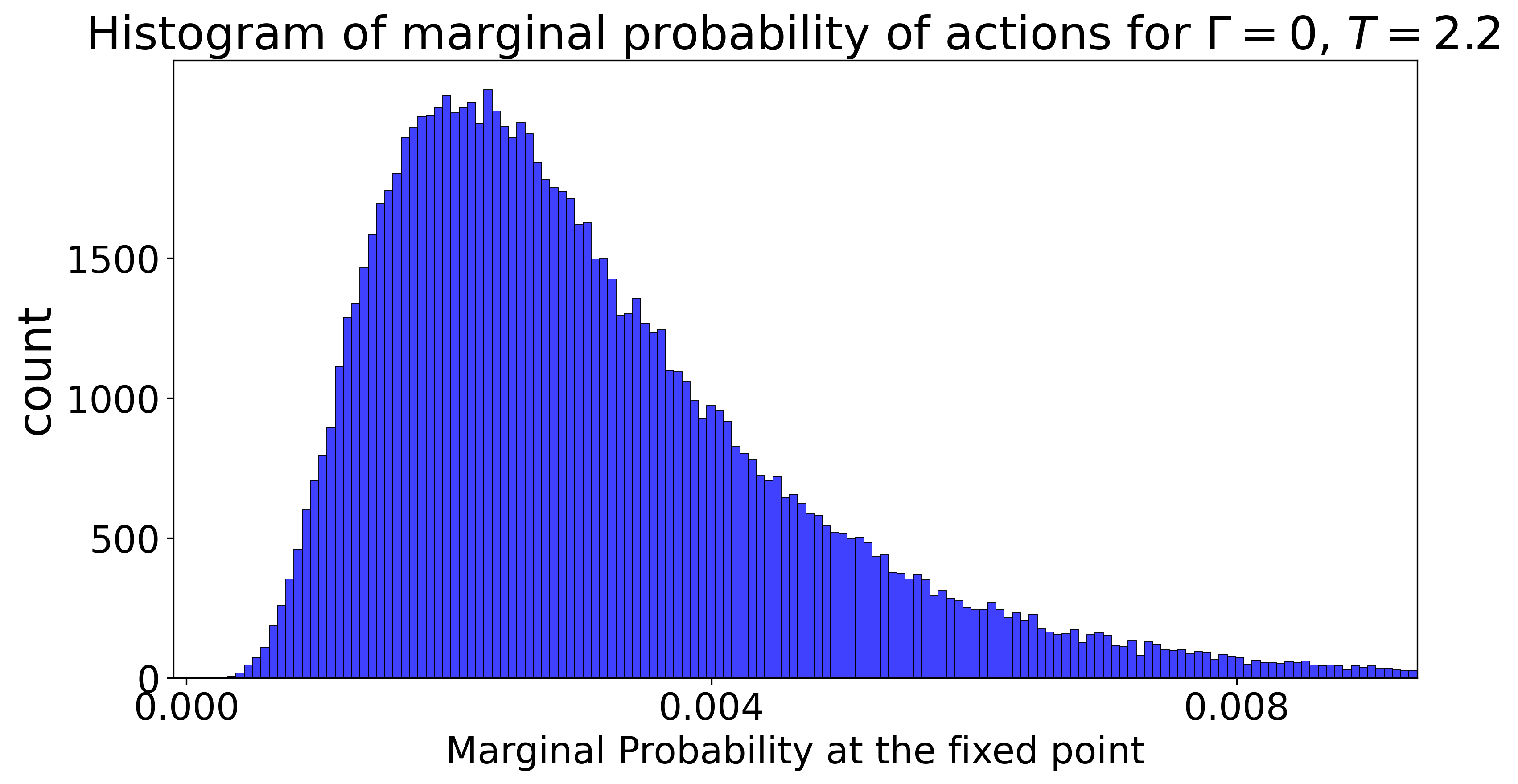}
        \label{fig:internal0}}
    \caption{Histogram of marginal probability of playing a random selected actions from ensemble numerical simulations.  These estimate the likelihood distribution of a randomly selected action under Q-Learning. For each plot, $160$ 2-player, $N=320$ action games are simulated with a randomly selected initial strategy until convergence.  The probabilities distribution of the actions at the fixed point is then recorded to form a histogram. On the left, the exploration rate $(T =1.79)$ remains in the unique fixed point regime, but is set below $\sqrt{3e/ 2} \approx 2.01 $ the threshold and a small fraction of strategies are played near 0.  On the right, the exploration rate $(T = 2.2)$ is above this threshold and there is no probability mass near 0.  This indicates the fixed points are internal.}
    \label{fig:figures_in_bound}
\end{figure*}

\begin{figure*}[h]
    \centering
    \subfigure{\centering
        \includegraphics[width=0.48\linewidth]{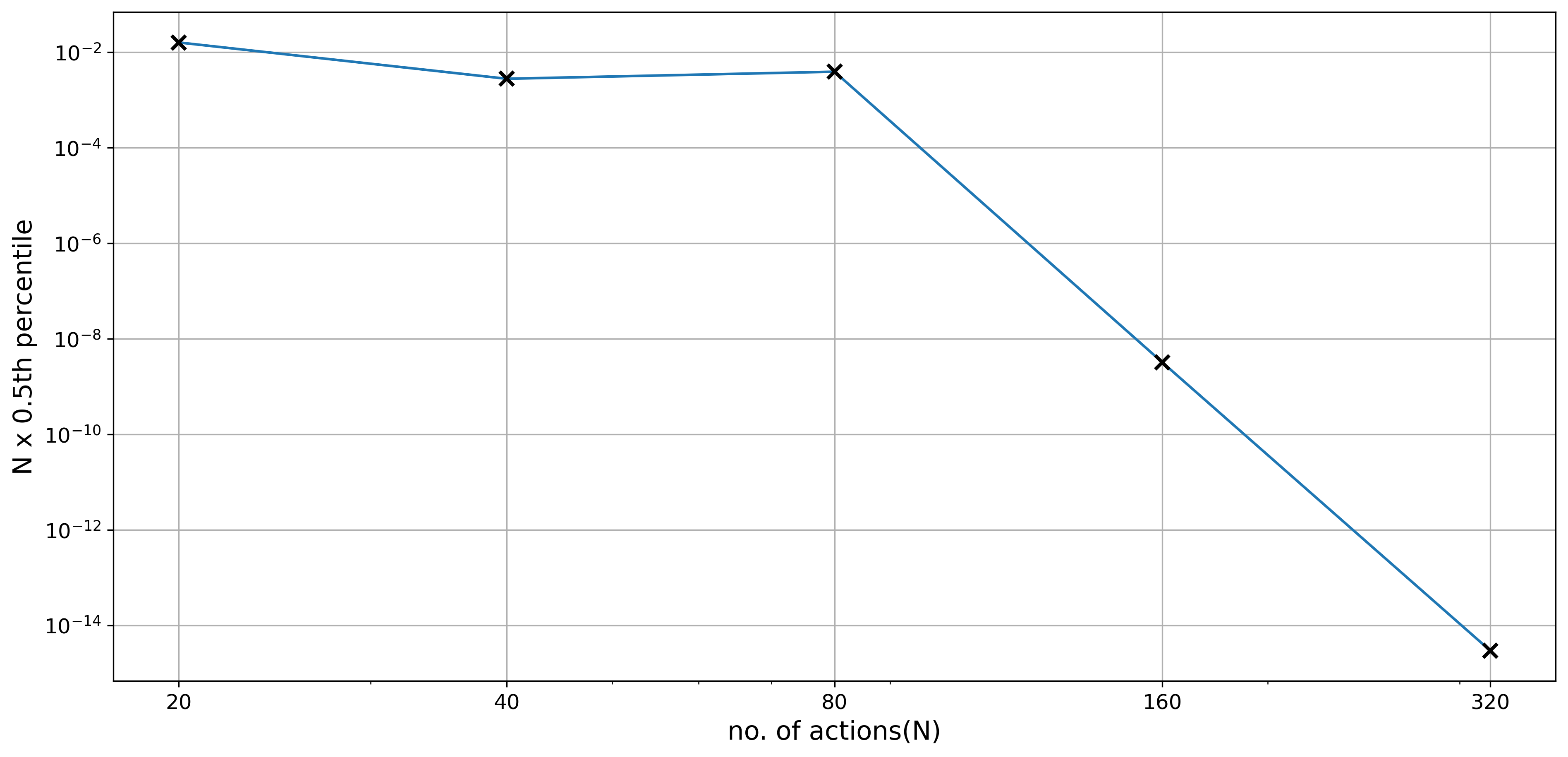}
        \label{fig:asymptotic_ex} 
    }
   \subfigure{\centering
        \includegraphics[width=0.48\linewidth]{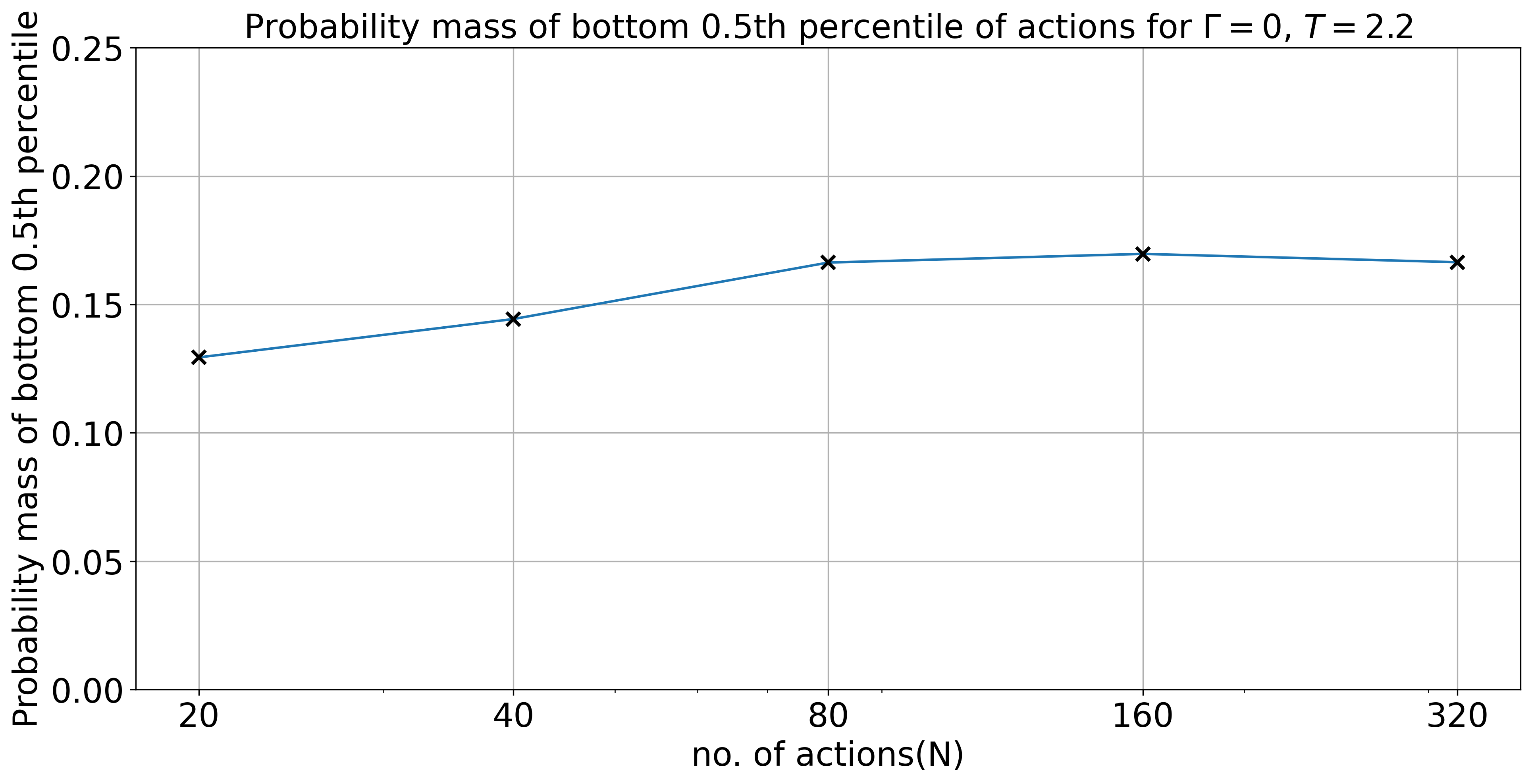}
        \label{fig:asymptotic_nex}}
    \caption{Plots of the how often the bottom 0.5th percentiles of action are played varying game size $N$ for games with varying exploration rate $T= 1.79, 2.2$.  For games with action sizes $N= [10, 20, 40, 80, 160 ,320]$ we run the following number of games: $[2560, 1280, 640, 320, 160]$ until convergence to obtain around 200,000 fixed points.
    When $T=1.79$ (left), we notice the bottom 0.5th percentiles of strategies are played with decreasing probability as the games scale. (Note the log scaling on the y-axes)  This corresponds to asymptotic extinction, where actions going extinct in the $N \to \infty$ limit.  When $T= 2.2 \geq \sqrt{3e(p-1)/ 2}$ (right), this does not occur. The probability mass of playing the bottom $x\%$ of actions does not depend on $N$ }
    \label{fig:figures_ex_nex}
\end{figure*}

Figures \ref{fig:figures_in_bound} demonstrate  when $\Gamma = 0$, the fixed points can be internal or on the boundary depending on the exploration rate $T$.  This is illustrated over sample 2-player games with varying $T$. The asymptotic extinction property of the internal fixed point is demonstrated in \ref{fig:figures_ex_nex}, which displays how the frequency of the playing a select $0.5\%$ of actions goes to zero as $N$ increases. \\

Solving the fixed point distribution of uncorrelated games varies from $\Gamma \neq 0$, due to the absence of the $a$ term.  The solve of this fixed point distribution will proceed in two parts:
\begin{itemize}
    \item[--] Showing the fixed points are internal when $T \geq \sqrt{3e(p-1)/ 2}$ and solving the fixed point distribution when $T \geq \sqrt{3e(p-1)/ 2}$
    \item[--] Solving the fixed point distribution when $T < \sqrt{3e(p-1)/ 2}$
\end{itemize}

\subsubsection{Fixed points are internal when $T \geq \sqrt{3e(p-1)/ 2}$}
To identify the critical exploration rate for fixed point to remain internal, we need to check the following distribution for $x(z)$ fulfils the self-consistency conditions \eqref{self_suff}:
\begin{align*} x(z) = Ke^{bz} , \forall z \in \mathbb{R} \end {align*}
The self-consistency condition corresponding to the $\chi$ term disappears.  We have to check the remaining to conditions.  A substitution of $x(z)$ yields:
\begin{align} 
q &= \int^{\infty}_{-\infty} x^2(z) Dz \notag \\ \notag
&= \frac{K}{\sqrt{2\pi}} \int^{\infty}_{-\infty} \exp\left( 2bz -\frac{z^2}{2} \right) dz \\ 
&= K \exp(2b^2)  \label{ss_eq_1}
\end{align}
\begin{align}
1 &= \int^{\infty}_{-\infty} x(z) Dz\notag  \\\notag
&= \frac{K}{\sqrt{2\pi}} \int^{\infty}_{-\infty} \exp\left( bz -\frac{z^2}{2} \right) dz \\
&= K \exp \left(\frac{b^2}{2}\right) \label{ss_eq_2}
\end{align}
Raising \eqref{ss_eq_2} to the $4^{th}$ power, and expression $q$ as a function of T we have:
\begin{align} \label{qq}
    q(T)= K^{-3}
\end{align}
Since there is no dependence on $a$, solving $q$ will yield us $b, K$ which are sufficient to find $x(z)$.
Substituting $b = \mathbf{q}^{(p-1)/2} T^{-1}$ and \eqref{qq} into \eqref{ss_eq_1}, we have:
\begin{align} \label{qt}
    T= \sqrt{\frac{3}{2} \left(\frac{\ln{(q(T))}}{q(T)^{(p-1)}} \right)}
\end{align}
It can easily be checked that \eqref{qt} has a solution for $q(T)$ when $T \geq \sqrt{3e(p-1)/ 2}$.  Thus, above this critical exploration rate, all fixed points for $\Gamma = 0$ are internal.
\subsubsection{Solving for $T < \sqrt{3e(p-1)/ 2}$}
For lower exploration rates, T, the fixed points are internal points.  $x(z)$ becomes:
\begin{align} \label{new_x}
x(z)= 
\begin{cases}
Ke^{bz} \; \; &, z < z_{\text{crit}} \\
0 \; \; &, z \geq z_{\text{crit}}
\end{cases}
\end{align}
Now, $z_{\text{crit}}$ becomes another parameter that has to be identified.  A substitution of \eqref{new_x} into the self-consistency relations yield:
\begin{align}  \label{ss_1}
q &= \int^{z_{\text{crit}}}_{-\infty} x^2(z) Dz \notag \\ \notag
&= \frac{K}{\sqrt{2\pi}} \int^{z_{\text{crit}}}_{-\infty} \exp\left( 2bz -\frac{z^2}{2} \right) dz \\ 
&= \frac{K}{2} \exp{\left(2b^2 \right) } \left( \text{erf} \left( \frac{\sqrt{2} (z_{\text{crit}} - 2b)}{2} \right) +1  \right) \notag \\
&= \frac{K}{2} \exp{(2b^2)} Q
\end{align}
\begin{align} \label{ss_2}
1 &= \int^{z_{\text{crit}}}_{-\infty} x(z) Dz \notag \\ \notag
&= \frac{K}{\sqrt{2\pi}} \int^{z_{\text{crit}}}_{-\infty} \exp\left( bz -\frac{z^2}{2} \right) dz \\ 
&= \frac{K}{2} \exp{\left(\frac{b^2}{2} \right) } \left( \text{erf} \left( \frac{\sqrt{2} (z_{\text{crit}} - b)}{2} \right) +1  \right) \notag \\
&= \frac{K}{2} \exp{\left(\frac{b^2}{2}\right)} E
\end{align}
where:
$$Q=  \text{erf} \left( \frac{\sqrt{2} (z_{\text{crit}} - 2b)}{2} \right) +1   $$
$$E = \text{erf }\left( \frac{\sqrt{2} (z_{\text{crit}} - b)}{2} \right) +1 $$
Substituting \eqref{ss_1} into \eqref{ss_2}. Noting $Q, E$ and $\mathbf{q}$ are dependent on $z_{\text{crit}}$ and $T$, we have the following expression for $\mathbf{q}$:
\begin{align} \label{new_q}
    q= 8K^{-3} (Q E^{-4})
\end{align}
A substitution into  \eqref{ss_1} yields.  The following relation:
\begin{align} 
q
=
\exp\left(\frac{3}{2} q^{(p-1)} T^{-2}\right) Q E^{-1}
\end{align}
$z_{\text{crit}}$ given by the maximum value of $z$ where there exists a solution for $q$ such that the above relation holds. This expression cannot be solved analytically, and has to be solved numerically. 
\subsection*{Derivation of the stability condition}
In this segment, we detail the stability analysis to obtain the following stability condition for the fixed points in the main paper: 
\begin{align*}\phi
\left\langle \bigg|
\frac{ T}{ x^\star} - \Gamma 
\mathbf{q}^{p-2} \chi
\bigg|^{-2}\right\rangle_{*} < \left((p-1) 
\mathbf{q}^{p-2} \right)^{-1}
\end{align*}
As discussed, the procedure will be broken up into the following steps:
\begin{itemize}
    \item[--] Introducing an $\varepsilon$-perturbation and linearising the dynamics at the fixed point.
    \item[--] Taking a Laplace Transform to work in the (complex) frequency domain.
    \item[--] Finding a stability criterion as a function of the Q-Learning parameters.
\end{itemize}
\subsubsection{Obtaining the Linearised Equations}
Beginning  with effective dynamics equation \eqref{eff_dyn2}, we introduce a small white noise term $\varepsilon(t)$ and perturb $x(t)$ and $\eta(t)$ slightly at their respective fixed points:
\begin{equation}\label{peturbation}
    x(t) = \mathbf{x^\star} + \underbrace{\hat{x}(t)}_{\mathcal{O}(\varepsilon)} \; \;,
    \; \; \eta(t) = \eta^\star + \underbrace{\hat{\eta}(t)}_{\mathcal{O}(\varepsilon)}
\end{equation}
Substituting \eqref{peturbation} into \eqref{eff_dyn2}, keeping only the $\mathcal{O}(\varepsilon)$ terms, we obtain the linearised equations about the fixed point:
\begin{equation}\label{linearisedeq}
\frac{d}{d t} \hat{x}(t) =  -T \hat{x}(t) + x^{\star} 
\left[ \Gamma \int \mathrm{d} t^{\prime} H(t, t') \hat{x}\left(t^{\prime}\right) + \hat{\eta}(t) + \varepsilon(t) \right]
\end{equation}
where:
\begin{itemize}
    \item[--] $H(t, t') = G\left(t, t^{\prime}\right) C\left(t, t^{\prime}\right)^{p-2}$
\end{itemize}
We note the following Taylor's Expansion result when substituting $x(t)$ into the $\ln$.
\begin{equation*}
    \ln(x(t)) = \ln(x^\star + \hat{x}) = \ln(x^\star) + \underbrace{\frac{\hat{x}}{x^\star}}_{\mathcal{O}(\varepsilon)} + \underbrace{\dots}_{\mathcal{O}(\varepsilon^2)}
\end{equation*}

\subsubsection{Laplace Transforms}
Now that we have obtained the linearised equations \eqref{linearisedeq}, the task is now to solve the following initial value problem: given an initial perturbation, predict the behaviour of $\hat{x}, \hat{\eta} \dots$ moving forward in time.  This is a linear initial value problem and thus, can be solved via the method of Laplace Transform.  The  Laplace transform is defined as follows:
\begin{equation*}
    \tilde{x}(\sigma)= \int^{\infty}_0\hat{x}(t) e^{-\sigma t} dt 
\end{equation*}
where:
\begin{itemize}
    \item[--] $\sigma \in \mathbb{C}$
\end{itemize}
The idea of the Laplace transform is to seek solution to \eqref{linearisedeq} expressed as a superposition of simpler solutions of the following form:
\begin{equation*}
    \hat{x}(t) = \tilde{x}(\sigma)e^{\sigma t}
\end{equation*}
where:
\begin{itemize}
    \item[--] $\sigma$ can be thought of the growth rate of the perturbation.  If $\Re(\sigma) > 0$, this perturbation grows in time. 
\end{itemize}
Taking the Laplace Transform of \eqref{linearisedeq}, we go from the time to the frequency domain, and through the following substitutions:
\begin{align*}
    &\hat{x}(t) \to \tilde{x}(\sigma) \; , \; \hat{\varepsilon}(t) \to \tilde{\varepsilon}(\sigma) \;,  \\
    &\hat{\eta}(t) \to \tilde{\eta}(\sigma) \; ,  \; H(t,t') \to \tilde{H}(\sigma)  \; , \; \frac{\partial}{\partial t } \to \sigma
\end{align*}
We obtain the following frequency relation:
\begin{equation}\label{modeq}
    A(\sigma, x^\star) \tilde{x}(\sigma) = \tilde{\eta}(\sigma) + \tilde{\varepsilon} (\sigma)
\end{equation}
where:
\begin{align*}
        A(\sigma, x^\star)= \frac{ \sigma + T}{x^\star} - \Gamma \tilde{H}(\sigma)
\end{align*}
A distinguishing feature between the Laplace Transform over the Fourier is its ability to handle complex eigenvalues, as it is a one-sided transformation.  Taking the square of \eqref{modeq} we have:
\begin{align}\label{squareeq}
 (\tilde{x}(\sigma))^2 =& \left( A(\sigma, x^\star) \right)^{-2} \times \left( (\tilde{\eta}(\sigma))^2 + (\tilde{\varepsilon}(\sigma))^2 + 2\tilde{\eta}(\sigma)\tilde{\varepsilon}(\sigma)  \right)
\end{align}
Integrating \eqref{squareeq} with respect to the all possible realisations we have:
\begin{align}\label{modeq2}
\left\langle |\tilde{x}(\sigma)|^2\right\rangle_{*} =& 
\left\langle | A(\sigma , x^{\star}) |^{-2}\right\rangle_{*} \times \phi
\left( \left\langle |\tilde{\eta}(\sigma) |^2\right\rangle_{*}  +
\left\langle |\tilde{\varepsilon}(\sigma)|^2\right\rangle_{*} \right)
\end{align}
where: 
\begin{itemize}
    \item[--] $\phi$ is the proportion of strategies which do not go extinct at the fixed point, which is given by $P(z < z_{\text{crit}}) , z \sim \mathcal{N}(0,1)$
\end{itemize}
We note:
\begin{itemize}
    \item[--] The $\tilde{\eta}(\sigma)\tilde{\varepsilon}(\sigma)$-term is eliminated as the expectation the product of random variable $\eta$ and an uncorrelated zero-mean variable $\varepsilon$ is 0.
    \item[--] $\phi$ takes into account only perturbations on the non-extinct strategies $x^{\star} > 0$ are relevant.  We suspect this scalar idea is that if a previously extinct strategy is re-introduced, it would go extinct again (see \citep{opper1992phase} for similar calculations).  We note $\phi = 1$ when $\Gamma < 0$, but $\phi < 1$ when $\Gamma > 0$.
\end{itemize}
\subsubsection{Finding a stability criterion}
We wish to now manipulate \eqref{modeq2} to generate a stability criterion.
From \eqref{eff_dyn2}, we make a substitution for $\tilde{\eta}$ term.  First note:
\begin{equation*}
    \left\langle \eta(t) \eta(t')\right\rangle_{*} =
    \left\langle x(t) x(t')\right\rangle_{*}^{p-1}
\end{equation*}
Subbing in \eqref{peturbation} and matching the $\mathcal{O}(\epsilon^2 )$ terms, we have:
\begin{align}
    \left\langle (\eta^{\star} + \hat{\eta}(t))^2  \right\rangle_{*} 
     &\to
    \left\langle |\tilde{\eta}(\sigma) |^2\right\rangle_{*}  \notag \\
    =\left\langle (\mathbf{x^{\star}} + \hat{x}(t))^2 \right\rangle_{*}^{p-1} 
     &\to (p-1) \left\langle (\mathbf{x^{\star}})^2 \right\rangle_{*}^{p-2} 
     \left\langle |\tilde{x}(\sigma) |^2\right\rangle_{*}
\end{align}
Substituting into \eqref{modeq2} and re-arranging we have:
\begin{equation}\label{modeq3}
\left\langle |\tilde{x}(\sigma) |^2 \right\rangle_{*}
=
\left\langle |\tilde{\varepsilon}(\sigma)|^2\right\rangle_{*} \notag \times
\left( \phi \left\langle |  A(\sigma , x^{\star}) |^{-2}\right\rangle_{*}^{-1}
-(p-1) \mathbf{q}^{p-2} 
\right)^{-1}
\end{equation}
Since the LHS is strictly positive, to avoid a contradiction, we need:
\begin{align}\label{contra_11}
\phi
\left\langle | A(\sigma , x^{\star}) |^{-2}\right\rangle_{*} < \left((p-1) 
\left\langle (\mathbf{x^{\star}})^2  \right\rangle_{*}^{p-2} \right)^{-1}
\end{align}
A fixed point becomes stable when the perturbation decays in time, this corresponds to $\Re\left(\frac{\partial}{\partial t} \right) = \Re(\sigma) < 0$.  The boundary of stability is given when $\Re(\sigma)  = 0$.  Thus, stability occurs where the following holds, $\forall k \in \mathbb{R}$:
\begin{align}\label{contra_12}
\phi
\left\langle \bigg|
\frac{ ik + T}{x^\star} - \Gamma 
\mathbf{q}^{p-2} \chi
\bigg|^{-2}\right\rangle_{*} < \left((p-1) 
\left\langle (\mathbf{x^{\star}})^2  \right\rangle_{*}^{p-2} \right)^{-1}
\end{align}
Taking a look at the LHS, we can split it into real and imaginary components.  We note it is maximised when $k=0$, when the imaginary component of the eigenmode $=0$, i.e:
\begin{align}\label{real_imag}\phi
\left\langle \bigg|
\underbrace{\frac{ ik}{x^\star}}_{\Im}
+ \underbrace{
\frac{T}{ x^{\star}}
- \Gamma 
\mathbf{q}^{p-2} \chi}_{\Re}
\bigg|^{-2}\right\rangle_{*} &= \phi
\left\langle \bigg| 
|\Im|^2 + |\Re|^2
\bigg|^{-1}\right\rangle_{*}\notag \\
& > \phi
\left\langle \bigg|
\frac{T}{ x^\star} - \Gamma 
\mathbf{q}^{p-2} \chi
\bigg|^{-2}\right\rangle_{*}
\end{align}
Thus, we obtain the following stability condition obtained in the main paper:
\begin{align*}\phi
\left\langle \bigg|
\frac{ T}{ x^\star} - \Gamma 
\mathbf{q}^{p-2} \chi
\bigg|^{-2}\right\rangle_{*} < \left((p-1) 
\mathbf{q}^{p-2} \right)^{-1}
\end{align*}
\subsubsection{A Discussion on the stability relation}
In the main paper, we mentioned there is a subtle difference between \citep{sanders2018prevalence} and our derivation.  This alongside, other previous work \citep{galla2013complex}  has used a Fourier Transform and only considered the real eigenmodes at 0, while we allow for complex eigenmodes. On the surface, the difference may appear somewhat irrelevant, as we recover the identical stability condition for $\Gamma < 0$. This raises the question: Why are we allowed to ignore complex eigenmodes when in transitions to instability? It seems than in numerical simulations holding a competitive game fixed $\Gamma <0$ and begining with a high exploration rate in the unique fixed point regime and lowering $T$, it appears system transitions to instability via a Hopf Bifurcation.  The formation of limit cycles also suggests complex eigenvalues are at play in destabilising Q-Learning. Further analysis is required to understand this effect.
\subsection*{Plots of selected simulations}
\begin{figure}[]
    \centering
    \begin{minipage}{\textwidth}
        \centering
        \includegraphics[height=2.2in]{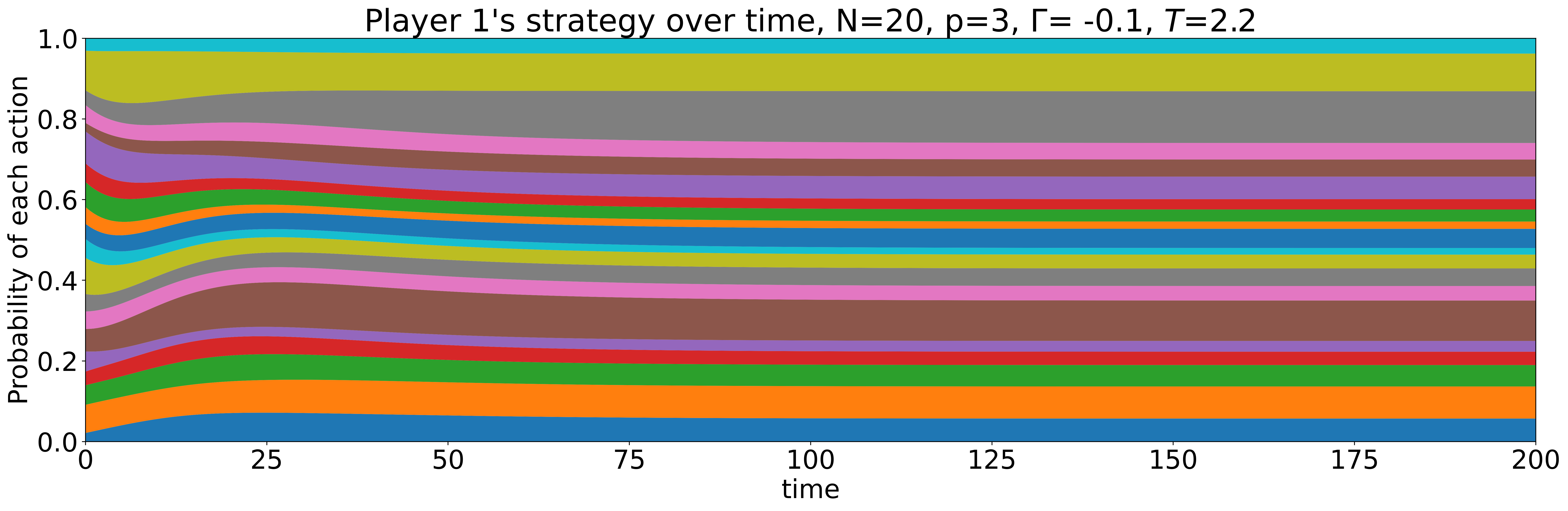}
    \end{minipage}

    \vspace{0.5cm} 

    \begin{minipage}{\textwidth}
        \centering
        \includegraphics[height=2.2in]{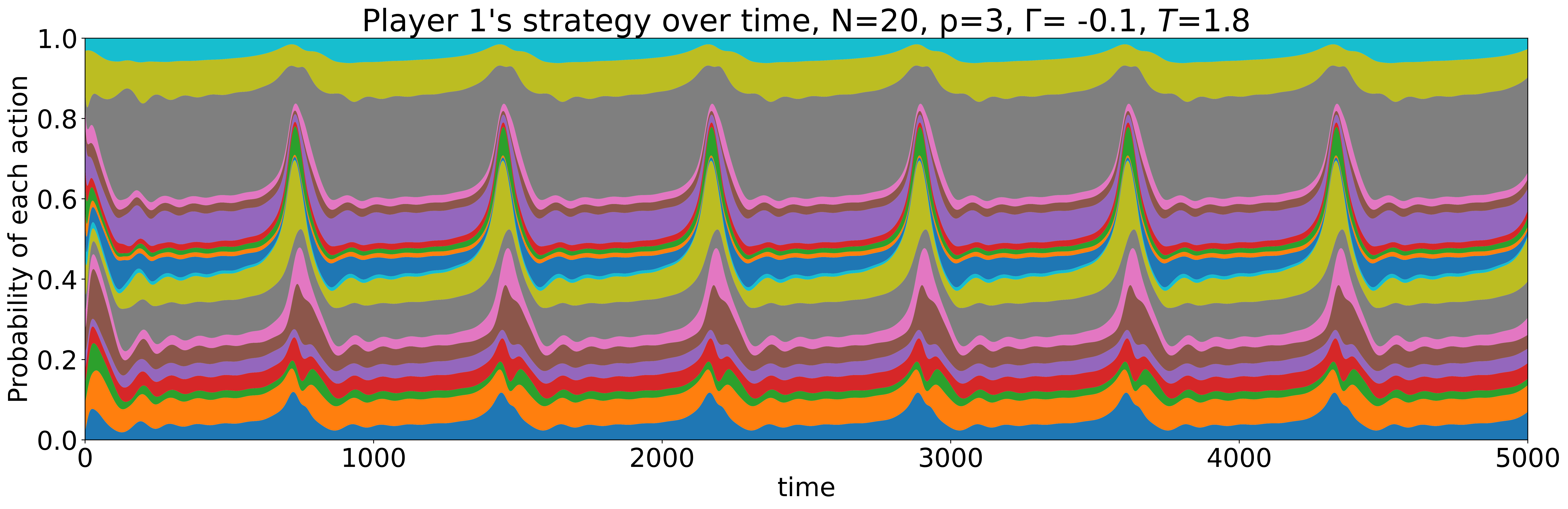}
    \end{minipage}

    \vspace{0.5cm} 

    \begin{minipage}{\textwidth}
        \centering
        \includegraphics[height=2.2in]{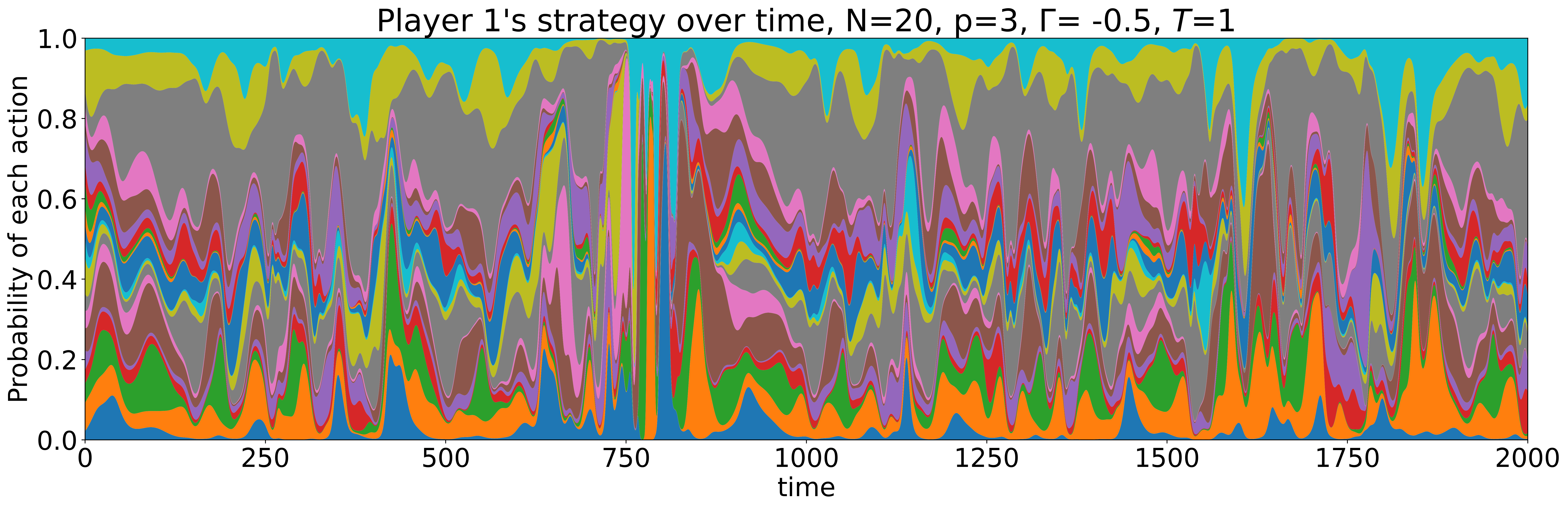}
    \end{minipage}
    \caption{Time series of strategy evolution under what appears to be different types of dynamical behaviour: convergence to a fixed point (top), a limit cycle (middle) and chaotic behaviour (bottom). At any given time, the probability over all actions must sum to 1.  Thus, we can represent strategy evolution with a stacked time chart, where each colour represents a unique action and the thickness represents the probability of the action being played.  The above plots demonstrate how the strategy of the first player evolves over time for different choices of exploration rate $T$ and $\Gamma$ in a $p=3$, $N=20$ action game.  Limit cycles occur occasionally on the boundary, but this highly depends on the initialisation of the payoffs.}
    \label{fig:time_series}
\end{figure}

\begin{figure*}[h]
    \centering
    \subfigure{\centering
        \includegraphics[width=0.48\linewidth]{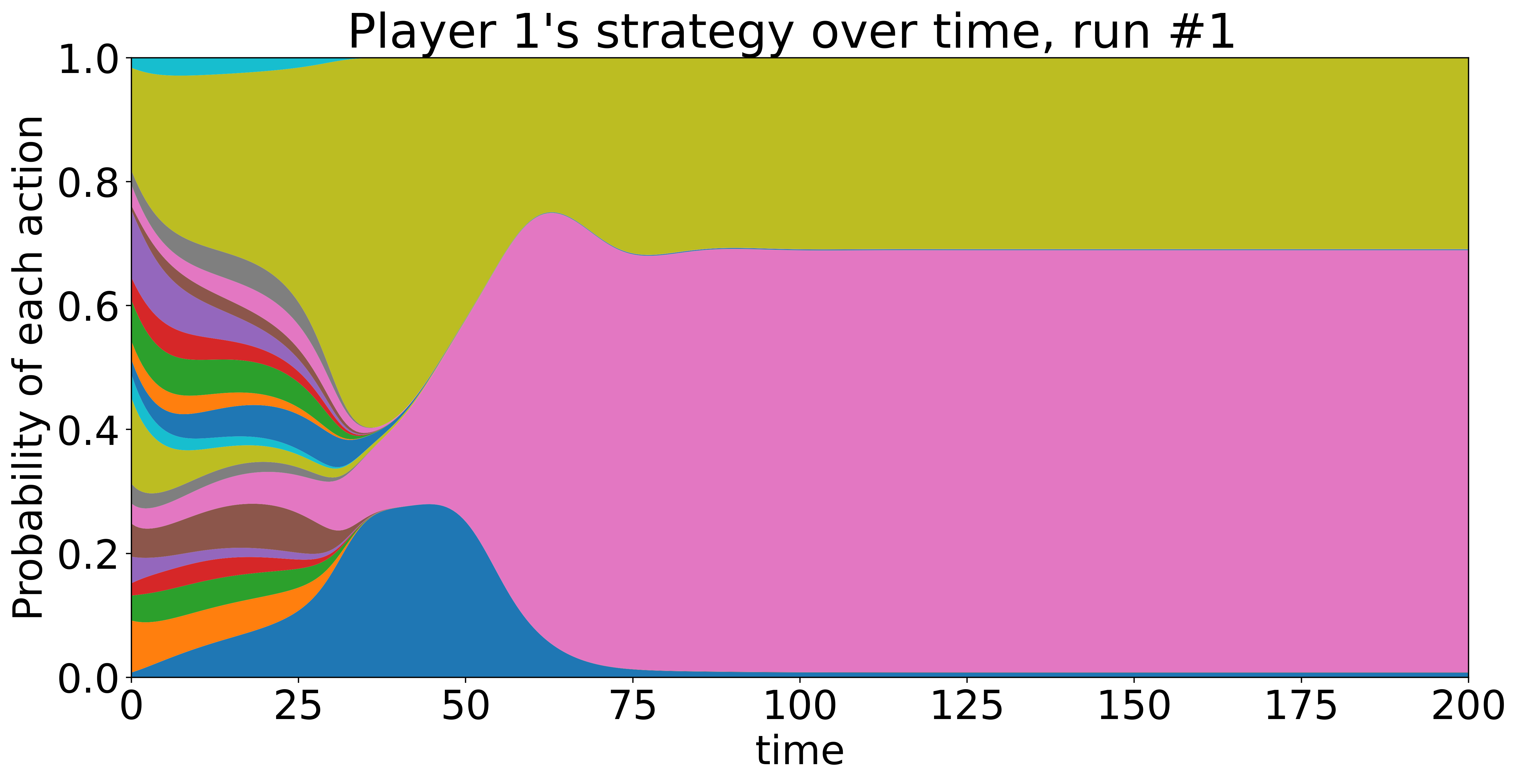}
        \label{fig:p3_extinction_1} 
    }
   \subfigure{\centering
        \includegraphics[width=0.48\linewidth]{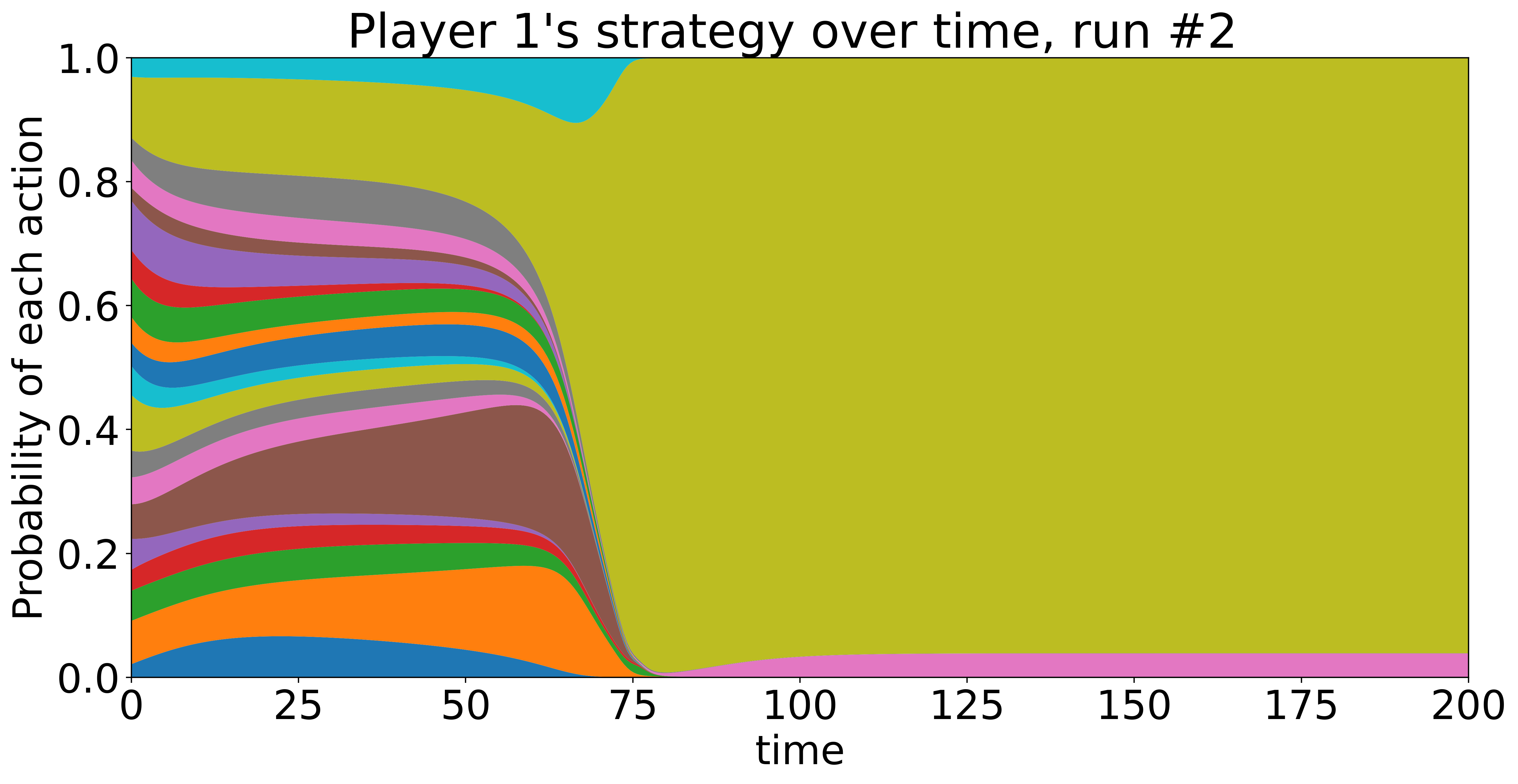}
        \label{fig:p3_extinction_2}}
    \caption{Convergence to different fixed points for two different initial conditions for the same randomly generated game of $p=3$, $N=20$ actions , when $\Gamma = 0.5$ , $T= 2.5$. }
    \label{fig:p3_extinction}
\end{figure*}

\begin{figure*}[h]
    \centering
    \subfigure{\centering
        \includegraphics[width=0.48\linewidth]{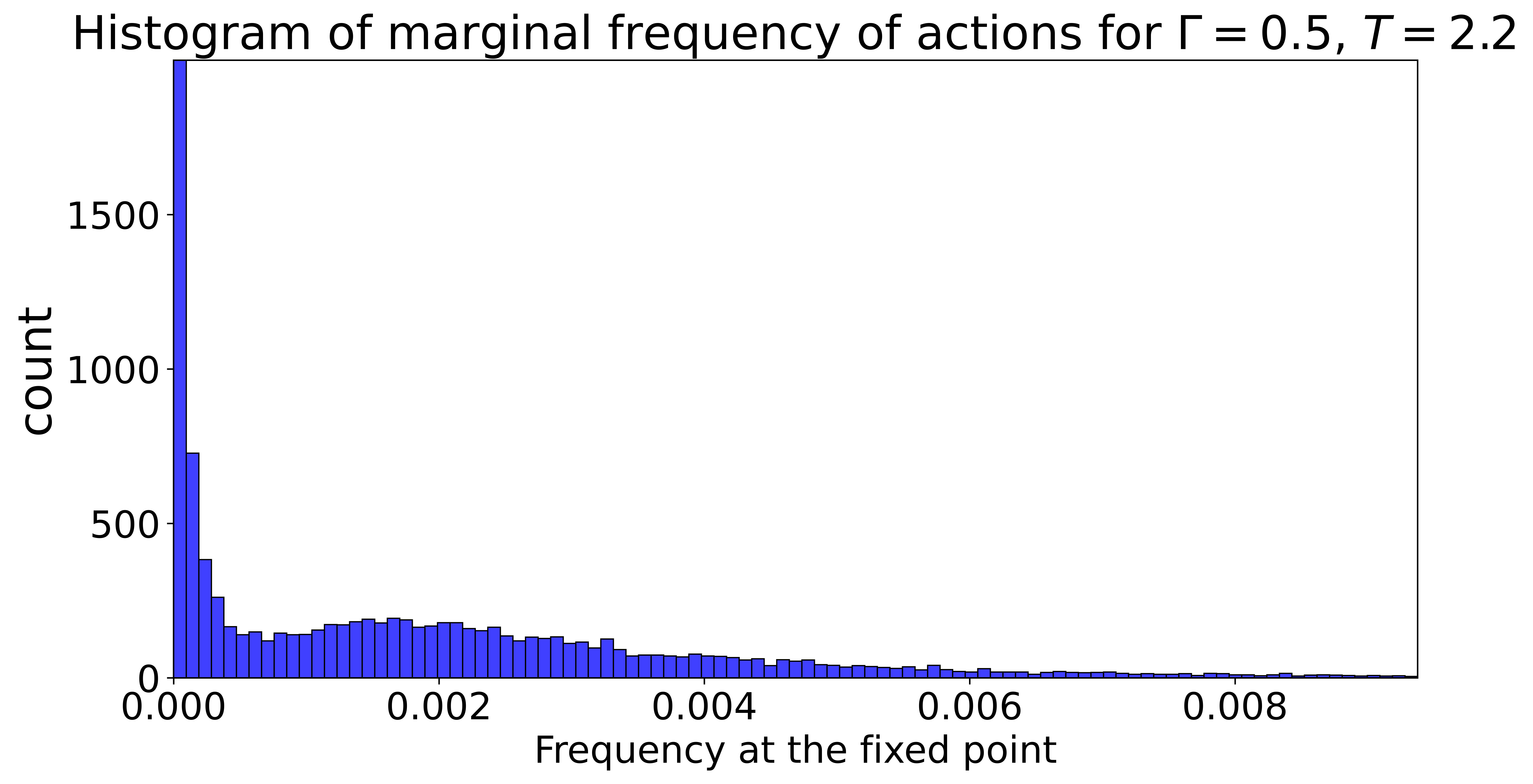}
        \label{fig:90_extinction_1} 
    }
   \subfigure{\centering
        \includegraphics[width=0.48\linewidth]{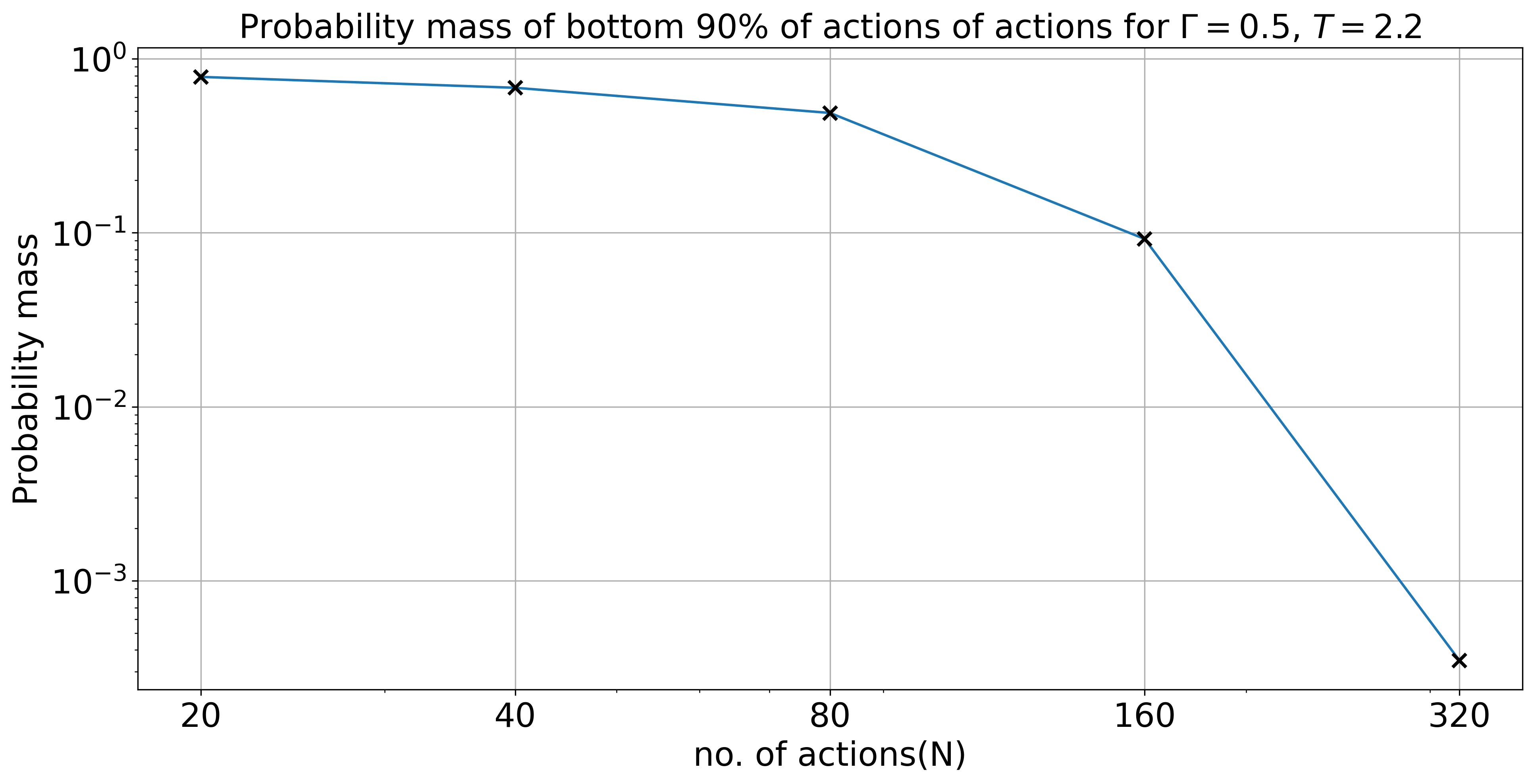}
        \label{fig:90_extinction_2}}
    \caption{ (Left) Histogram plot of showing the marginal distribution of how often actions are being played from  ensemble $N=320$-action 2-player games $\Gamma= 0.5, T = 2.2$ at the fixed point from random initial conditions.   This corresponds to the multiple fixed point regime. Note the concentration of strategies played with near 0 probability. (Right) Holding the rest of the parameters fixed and varying $N$, a plot of the probability mass of the bottom $90\%$. actions is generated.  This implies there is a concentration of probabilities on a small fraction of actions at low exploration rates in $\Gamma > 0$.}
    \label{fig:90_extinction}
\end{figure*}

\begin{figure*}[h]
    \centering
    \subfigure{\centering
        \includegraphics[width=0.48\linewidth]{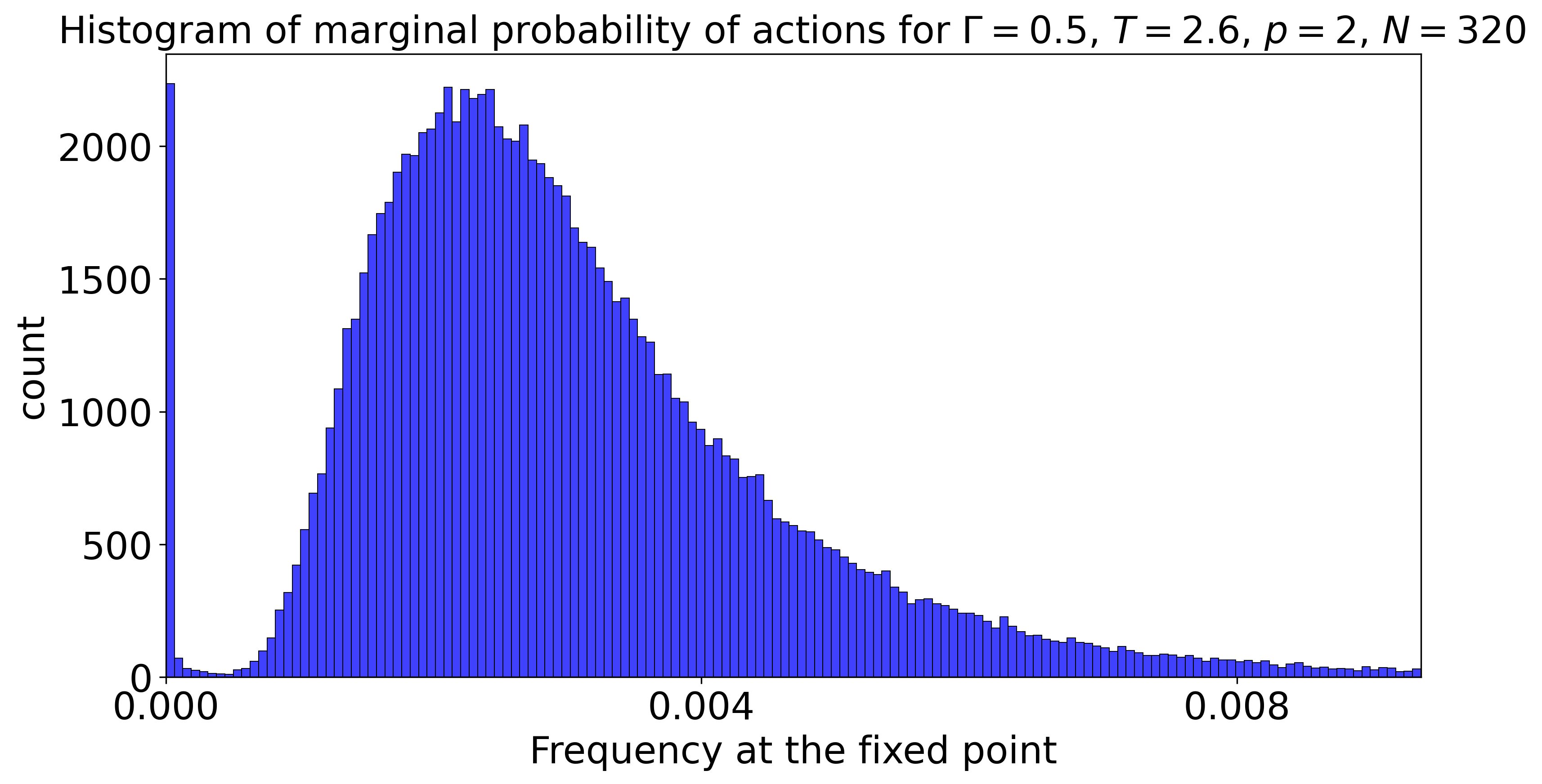}
        \label{fig:no_contra} 
    }
   \subfigure{\centering
        \includegraphics[width=0.48\linewidth]{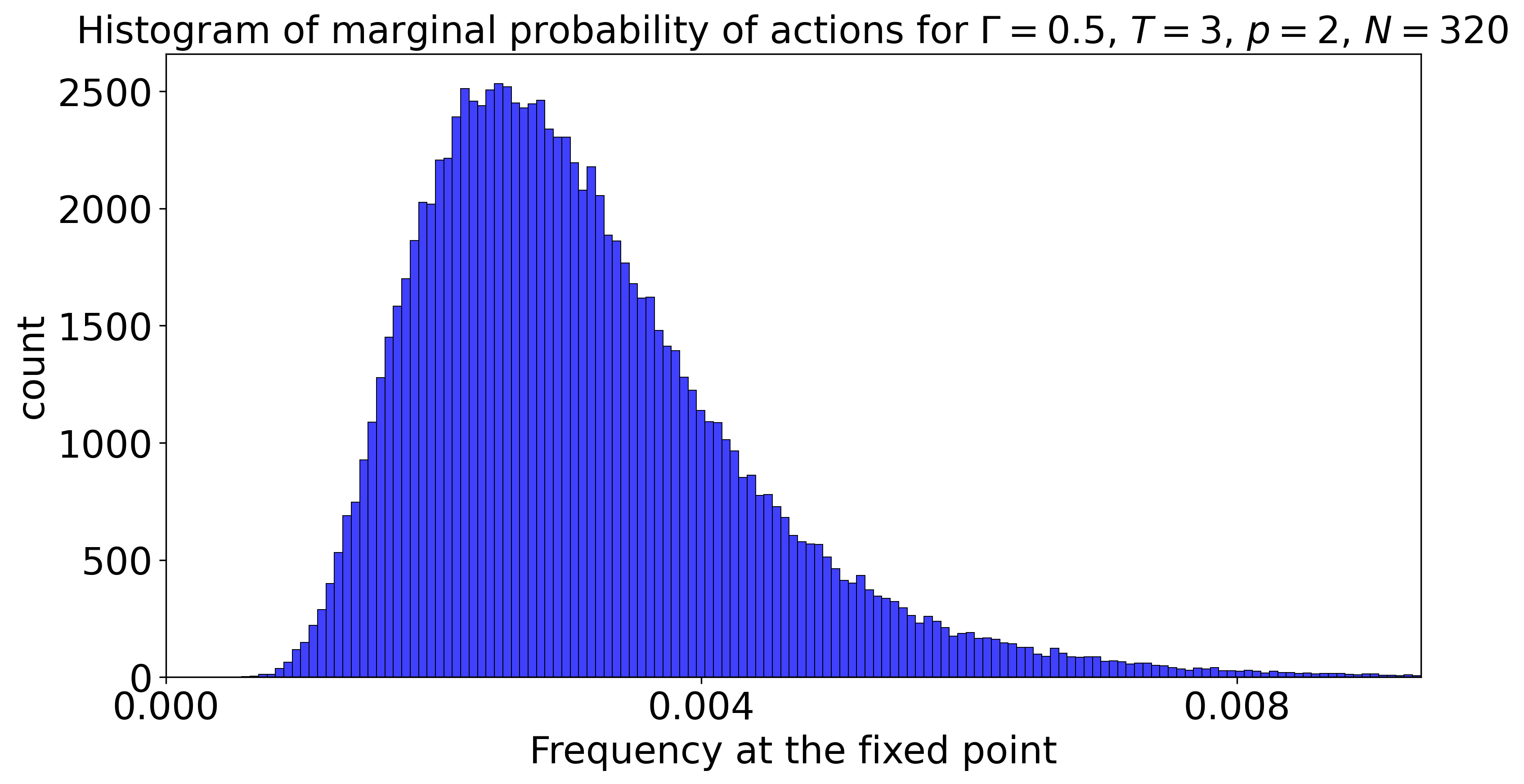}
        \label{fig:apparent_contra}}
    \caption{  Histogram plot of showing the marginal distribution of how often actions are being played from  160 random trajectories from $N=320$-action 2-player games $\Gamma= 0.5$ for $T = 2.6$ (left) and $T=3$ (right).  Both these parameter are in the unique fixed point regime given by $T > T_{\text{crit}} = 2.51$.  It appears strategies only appear to go extinct on the left plot ($T= 2.6$).  As discussed in the main paper, this is likely due to extinctions being extremely rare for parameters away from the boundary.  At $T=3$, extinctions are expected to be approximately a  1 in $10^8$ event, making it difficult to verify.}
    \label{fig:contradiction?}
\end{figure*}

\end{document}